\documentclass[reprint,
superscriptaddress,
preprintnumbers,
nofootinbib,
nobibnotes,
amsmath,amssymb,
aps, prd,
]{revtex4-2}

\usepackage[pdftex]{xcolor}
\usepackage{dsfont}
\usepackage{graphicx}
\usepackage{dcolumn}
\usepackage{bm}
\usepackage{hyperref}

\usepackage{tikz}
\usetikzlibrary{decorations.markings}
\usetikzlibrary{shapes.geometric, arrows, calc, math}

\newcommand{\phizero}{\phi^{(0)}}
\newcommand{\phin}{\phi^{(n)}} 
\newcommand{\phinpo}{\phi^{(n+1)}} 
\newcommand{\osub}{^{(0)}}
\newcommand{\nsub}{^{(n)}}
\newcommand{\nposub}{^{(n+1)}}

\newcommand{\argmin}{\ensuremath{\mathop{\rm arg\;min}}}
\newcommand{\argmax}{\ensuremath{\mathop{\rm arg\;max}}}

\newcommand{\tr}{\ensuremath{\mathop{\rm tr}}}

\newcommand{\bv}[1]{\bm{#1}}

\begin{document}

\preprint{TUM-EFT 202/25}
\preprint{FERMILAB-PUB-0041/T}
\preprint{MIT-CTP/5995}
\doi{10.1103/qjjp-qw7x}

\title{Wilson loops with neural networks}

\author{Verena Bellscheidt}
\email{verena\_b@mit.edu}
\affiliation{Center for Theoretical Physics---A Leinweber Institute,
Massachusetts Institute of Technology, 77~Massachusetts Avenue, Cambridge, Massachusetts 02139, USA}

\author{Nora Brambilla}
\email{nora.brambilla@ph.tum.de}
\affiliation{Department of Physics, School of Natural Sciences, Technische Universit\"at M\"unchen, James-Franck-Strasse~1, 85748 Garching, Germany}
\affiliation{Institute for Advanced Study, Technische Universit\"at M\"unchen,
Lichtenbergstrasse~2a, 85748 Garching, Germany}
\affiliation{Munich Data Science Institute, Technische Universit\"at M\"unchen, \\
Walther-von-Dyck-Strasse~10, 85748 Garching, Germany}

\author{Andreas~S.~Kronfeld}
\email{ask@fnal.gov}
\affiliation{Theory Division, Fermi National Accelerator Laboratory,
P.O.~Box~500, Batavia, Illinois 60510, USA}
\affiliation{Institute for Advanced Study, Technische Universit\"at M\"unchen,
Lichtenbergstrasse~2a, 85748 Garching, Germany}

\author{Julian Mayer-Steudte}
\email{julian.mayer-steudte@tum.de}
\thanks{Contact author}
\affiliation{Department of Physics, School of Natural Sciences, Technische Universit\"at M\"unchen, James-Franck-Strasse~1, 85748 Garching, Germany}
\affiliation{Munich Data Science Institute, Technische Universit\"at M\"unchen, \\
Walther-von-Dyck-Strasse~10, 85748 Garching, Germany}

\collaboration{TUMQCD Collaboration}\noaffiliation

\date{1 May 2026}

\begin{abstract}
Wilson loops are essential objects in QCD and have been pivotal in scale setting and demonstrating confinement.
Various generalizations are crucial for computations needed in effective field theories.
In lattice gauge theory, Wilson loop calculations face challenges, including excited-state contamination at short times and the signal-to-noise ratio issue at longer times.
To address these problems, we develop a new method by using neural networks to parametrize interpolators for the static quark-antiquark pair.
We construct gauge-equivariant layers for the network and train it to find the ground state of the system.
The trained network itself is then treated as our new observable for the inference.
Our results demonstrate a significant improvement in the signal compared to traditional Wilson loops, performing as well as Coulomb-gauge Wilson-line correlators while maintaining gauge invariance.
Additionally, we present an example where the optimized ground state is used to measure the static force directly, as well as another example combining this method with the multilevel algorithm.
Finally, we extend the formalism to find excited-state interpolators for static quark-antiquark systems.
To our knowledge, this work is the first study of neural networks with a physically motivated loss function for Wilson loops.
\end{abstract}

\maketitle

\tableofcontents

\section{Introduction}

The Wilson loop~\cite{Wilson:1974sk} is a widely used tool in mathematical and computational studies of gauge theories.
In particular, in confining theories, such as quantum chromodynamics (QCD), Wilson loops of various shapes yield information about the energy carried by gluons in the presence of static sources.
Energies are, of course, spectral quantities, which depend crucially on some properties of the loop(s) while being independent of other features.

This circumstance can be illustrated by a rectangular Wilson loop of spatial separation $\bv{r}$ and temporal extent $t$.
For large (Euclidean)~$t$, a spectral decomposition shows the Wilson loop to be a sum of terms of the form
\begin{equation}
    \sum_n |c_n|^2\, e^{-taE_n(\bv{r})} ,
    \label{eq:En}
\end{equation}
where the ``static energies'' $E_n(\bv{r})$ depend on the separation (usually just on the distance $r=|\bv{r}|$) and the representation in which the Wilson loop is evaluated.
In the lattice-gauge-theory literature and elsewhere, the static energies are usually referred to as ``static potentials.''
The leading static energy $E_0(r)$ is the quarkonium  static potential\footnote{However, in the perturbative regime of small~$r$, the potential suffers from infrared divergences~\cite{Appelquist:1977es} that are canceled in the static energies by contributions from ultrasoft gluons that can be identified in a multipole expansion~\cite{Brambilla:1999qa,Brambilla:1999xf}.  Nonperturbatively, the static energy coincides with the static potential.}
Higher excitations of the static energies $E_n(r)$ correspond to the static potentials of hybrid states (with valence gluons) and, in the presence of light quarks, of tetraquark states~\cite{Brambilla:1999xf, Brambilla:2004jw,Berwein:2024ztx}.

As spectral quantities, the energies in Eq.~\eqref{eq:En} remain the same if the spatial parts of the Wilson loop are replaced by any spatial path connecting the same end points.
Instead, the coefficients $c_n$ depend on the spatial paths.
In fact, it is possible and advantageous to design a combination of paths to isolate a state of particular interest.

For example, in lattice gauge theory, a common strategy is to fix the lattice gauge field to Coulomb gauge and then take the gauge-field average of Wilson lines; see, for example, Ref.~\cite{Brambilla:2022het}.
Gauge averaging selects out combinations of paths that are gauge invariant, so the original gauge-invariant static energies are obtained.
The static energies are presumably unaffected by the Gribov ambiguity of the Coulomb gauge.
That said, experience~\cite{Brambilla:2022het} has shown that the best practice is to use the same algorithm with the same tolerances and stopping conditions for all configurations in an ensemble.
Otherwise, somewhat different combinations of paths are implicitly chosen.

Such problems can be avoided with an explicitly gauge-invariant methodology that smears the spatial path, thereby maintaining the main virtue of the Coulomb gauge.
In this paper, we use neural networks to generate an optimal set of paths.
We develop and test the neural network on the lowest-lying static energy, finding that the neural network performs as well as the Coulomb gauge over a wide range of distances.
We then adapt the neural network to optimize for the first excitation, which has different rotational quantum numbers.
This neural network finds two degenerate levels---as expected. 

A~further advantage of the neural-network approach is that it can be applied to more complicated Wilson loops, once the basic setup has been established.
Here, we proceed to develop and test a neural network for the \emph{static force}, which is the spatial derivative of the static energy.
We compute it not by an explicit discrete difference but by the insertion of plaquettes along the temporal Wilson lines~\cite{Vairo:2016pxb,Brambilla:2021wqs,Brambilla:2023fsi}.
Many other examples of modified Wilson loops~\cite{Eichten:1980mw,deForcrand:1985zc,Campostrini:1986ki, Barchielli:1986zs,Brambilla:2000gk,Pineda:2000sz,Berwein:2015vca,Capitani:2018rox,Soto:2020xpm,Schlosser:2021wnr,Eichberg:2024svw, Berwein:2024ztx} should be amenable to the neural-network technique, which we leave for future work.

While earlier works have applied neural networks to Wilson loops~\cite{Favoni:2020reg, Nagai:2025rok}, this is---to our knowledge---the first to introduce a physically motivated loss function.
Complementary machine-learning-based approaches to improve the operators and the signal are contour deformation of the path integral~\cite{Detmold:2020ncp,Detmold:2021ulb,Lin:2023svo} and renormalization-group-improved actions~\cite{Holland:2025fsa}.
For reviews of further work on machine learning in lattice gauge theory, see Refs.~\cite{Kanwar:2024ujc,Tomiya:2025quf}.

The rest of this paper is organized as follows.
Section~\ref{sec:theo} presents the notation and theoretical background for Wilson loop studies in lattice gauge theory.
The building blocks for neural networks are discussed in Sec.~\ref{sec:NN}.
Results for the lowest-lying static energy, a matrix element for the static force, and excited states are presented in Sec.~\ref{sec:results}.
Section~\ref{sec:outlook} concludes with a few remarks on future applications.

\section{Theoretical background}
\label{sec:theo}

\subsection{Wilson loops}

A rectangular, on-axis Wilson loop is defined in terms of link variables as
\begin{equation}
    W_{r\times t} = S(\bv{x},\bv{x}+\bv{r},0)T(0,t,\bv{x}+\bv{r}) 
        S(\bv{x}+\bv{r},\bv{x},t)T(t,0,\bv{x}) ,
    \label{eq:Wilson_loop_definition}
\end{equation}
where the factors are
\begin{align}
    S(\bv{x},\bv{x}+\bv{r},t) &= \Pi_{k=0}^{r-1}U_r(\bv{x}+k\hat{\mathbf{e}}_r,t)\label{eq:spatial_wilson_line_forward}\\
    S(\bv{x}+\bv{r},\bv{x},t) &= S(\bv{x},\bv{x}+\bv{r},t)^\dagger\label{eq:spatial_wilson_line_backward}\\
    T(0,t,\bv{x}) &= \Pi_{k=0}^{t-1}U_4(\bv{x},k)\label{eq:temporal_wilson_line_forward}\\
    T(t,0,\bv{x}) &= T(0,t,\bv{x})^\dagger
    \label{eq:temporal_wilson_line_backward}
\end{align}
where $\hat{\mathbf{e}}_r$ is a unit vector in the direction of $r$, and $U_\mu(x)=U_\mu(\bv{x},t)$ are the gauge links in direction $\mu=1,2,3,4$ ($4$ is the Euclidean time direction) at lattice site $x=(\bv{x},t)$, $\bv{x}=(x_1,x_2,x_3)$. The Wilson loop starts and ends at $\bv{x}$ in the time slice $t=0$ with residing in the $\hat{\mathbf{e}}_r$--$\hat{\mathbf{e}}_4$ plane with spatial and temporal extension $r$ and $t$, respectively. Here, and for the rest of the paper, $r$ and $t$ are stated in lattice units, i.e., $t_\mathrm{ph}=ta$ and $r_\mathrm{ph}=ra$, where $a$ is the lattice spacing.

A gauge-invariant quantity is obtained by taking the trace of the Wilson loop.
We denote the gauge-field average as $\langle \tr W_{r\times t}\rangle$.
In a numerical simulation, it is approximated with a finite lattice ensemble as
\begin{align}
    \langle \tr W_{r\times t}\rangle \approx \frac{1}{N_\mathrm{conf}}\sum_{i=1}^{N_\mathrm{conf}}\tr W_{r\times t}^{(i)}
\end{align}
where the $(i)$ superscript indicates the measurement of the observable on the $i$th configuration.
Additionally, averaging is performed over all origins of the loop and all three spatial directions for $\mathbf{e}_r$.

The trace yields a globally gauge-invariant object, but individual parts are gauge variant.
For example, imposing a gauge transformation $G(x)$, such that the gauge field transforms as
\begin{align}
    U_\mu(x) \rightarrow G(x)U_\mu(x)G^\dagger(x+\hat{e}_\mu) ,
\end{align}
transforms the spatial Wilson lines accordingly to
\begin{align}
    S(\bv{x},\bv{x}+\bv{r},t) \rightarrow G(\bv{x},t)S(\bv{x},\bv{x}+\bv{r},t) G^\dagger(\bv{x}+\bv{r},t),
\end{align}
which corresponds to the gauge property of a static quark-antiquark pair located at $(\bv{x},t)$ and $(\bv{x}+\bv{r},t)$.

The Wilson loop is interpreted as a two-point function of a mesonic state, $|S_r\rangle$, composed of two static quarks separated by $r$ where $S(\bv{x}+\bv{r},\bv{x},t)\cong S_r(t)$ is its interpolator.
Therefore, the Wilson loops represent transfer matrix elements of $S(\bv{x}+\bv{r},\bv{x},t)$ as
\begin{align}
    \langle \tr W_{r\times t}\rangle \cong \langle S_r(0)S_r^\dagger(t)\rangle = \langle S_r|T^{t}|S_r\rangle
    \label{eq:transer_matrix}
\end{align}
with the transfer matrix $T=e^{-aH}$.
Assuming positivity, or at least physical positivity~\cite{Luscher:1984is}, of the transfer matrix, the spectral decomposition of the Wilson loop is
\begin{align}
    \langle \tr W_{r\times t}\rangle = \sum_{n=0}^\infty |c_n|^2e^{-taE_n(r)}
    \label{eq:spectral_representation_Wilson_loop}
\end{align}
where $c_n=\langle n|S_r\rangle$ is the overlap of state $n$ with the spatial Wilson line and $E_n(r)$ the energy (in physical units, hence the lattice spacing $a$) of the $n$th state.
The ground-state energy $E_0(r)$, i.e., $E_0(r)<E_n(r)$ for $n>0$, defines the static energy of the heavy quark-antiquark pair, and it can be used in a Schr\"odinger equation to calculate the energy levels of quarkonium, hybrids, or tetraquarks~\cite{Brambilla:1999xf, Brambilla:2004jw, Berwein:2024ztx}.

\subsection{Generalized Wilson loops}

The Wilson loop Eq.~\eqref{eq:Wilson_loop_definition} is one single, closed loop of link variables; hence, $W_{r\times t}\in \mathrm{SU}(3)$.
The definition of the Wilson loop generalizes by considering Wilson loops with field-strength component insertions in the temporal 
or spatial Wilson lines, replacing the spatial Wilson lines with sums of paths with different shapes, inserting light quark 
pairs in the spatial Wilson lines, or a combination of those possibilities, or both.
For example, the spatial Wilson lines can be replaced with sums over paths of varying shapes, which can then project better onto specific states $|n\rangle$.

Inserting one chromoelectric field component on the temporal Wilson line yields the ground-state matrix element of the chromoelectric field component, which is a measure for the static force between the static quark-antiquark pair~\cite{Vairo:2016pxb,Brambilla:2021wqs,Brambilla:2023fsi}.
Multiple insertions of chromoelectric and chromomagnetic field components in the temporal Wilson lines yield the spin and relativistic correction terms to the quarkonium static potential~\cite{Eichten:1980mw,deForcrand:1985zc, Campostrini:1986ki,Barchielli:1986zs,Bali:1997am,Brambilla:2000gk,Pineda:2000sz,Brambilla:2004jw}.

Insertions of chromomagnetic and chromoelectric field components and their combinations in the spatial Wilson line produce interpolators with a specific quantum number, creating the hybrid states. These interpolators are not unique and are replaced by sums of differently shaped paths that still obey the correct quantum number but have better overlap with the hybrid state in question.
These ideas have been extensively studied in the literature~\cite{Juge:1997nc,Juge:1999ie,Juge:2002br,Capitani:2018rox, Schlosser:2021wnr,Hollwieser:2025nvv}.
Insertions of chromoelectric and chromomagnetic field components in the temporal Wilson lines together with insertion of appropriate interpolators in the spatial lines yield spin and relativistic correction terms to the hybrids and tetraquark static potentials~\cite{Berwein:2015vca,Soto:2020xpm,Berwein:2024ztx}.

\subsection{Neural networks for Wilson loops}
\label{sec:nnet_wilson_loop_intro}

In this work, we study the description of Wilson loops with a neural network designed to parametrize optimal interpolators with respect to the ground-state overlap.
We propose replacing the spatial Wilson line by a neural network $\tilde{S}$ with the same end points as $S$.
At the same time, we keep the temporal Wilson lines fixed.
The neural network $\tilde{S}$ depends on a set of parameters $\{w\nsub\}$, where $n$ iterates through the layers of the network.
By optimizing these parameters, we maximize the ground-state overlap $|c_0|^2$ while minimizing the excited-state contributions $|c_{n}|^2$, $n>0$.
By leaving the temporal Wilson lines untouched, we preserve the spectrum $E_n(r)$.

To retain gauge invariance, the gauge-transformation property of $\tilde{S}$ has to be 
the same as for $S$; i.e., under a gauge transformation $G(x)$, it must transform as
\begin{align}
    \tilde{S}(\bv{x},\bv{x}+\bv{r},t)\rightarrow G(\bv{x},t)\tilde{S}(\bv{x},\bv{x}+\bv{r},t)G^\dagger(\bv{x}+\bv{r},t).
\end{align}
We achieve this by starting with an initial set of paths that obey the correct gauge property and recombining them by iterating through layers of a neural network.

A useful network has to fulfill several requirements, stemming from gauge invariance and practical considerations.
First, the layers must be gauge equivariant, meaning that the output elements conform to the same gauge property as the input elements.
Second, the layers must exhibit high expressivity in representing the underlying gluon dynamics, and the network must systematically develop complicated structures.
Third, the input elements and layers should be simple to implement and to generalize to different separations $r$ and directions
$\mathbf{e}_r$, thereby avoiding the explicit construction of complicatedly shaped objects.
The new, generalized Wilson loop $\widetilde{W}_{r\times t}(\{ w\nsub\})$ is now parametrized by a set of parameters
$\{w\nsub\}$ and can be optimized with respect to a loss function $L$.
The fourth and final requirement is that the derivative $\partial L/\partial w\nsub$ must exist.
These four requirements will be resolved in this study by developing and implementing gauge-equivariant layers, imposing a loss function,
training the neural network, and demonstrating its capability.

\section{Building blocks}
\label{sec:NN}

This section covers the building blocks of our neural-network approach. We cover the definition of gauge-equivariant layers, the general neural-network architecture, and the process of determining the loss function. Those technical details are crucial for establishing the foundation for further applications of the neural-network method.

\subsection{Layers for general states}
\label{sec:layers_general}

To parametrize general static quark-antiquark states, we employ a feedforward neural network whose layers iteratively recombine open paths sharing the gauge transformation law of the straight spatial Wilson line.
As emphasized in Sec.~\ref{sec:nnet_wilson_loop_intro}, the guiding principle is gauge equivariance: If each layer maps inputs with the static-line gauge property to outputs with the same property, the full network inherits this equivariance.
Specifically, we restrict ourselves to algebraic operations---most notably linear superpositions and simple local compositions---that preserve the end points of the paths and thus their gauge behavior, ensuring that the network output remains a valid interpolator for the static pair.
The trainable weights are chosen to be identical for all lattice sites and all orientations of $\bv{r}$, thereby encoding translational and rotational (on the cubic lattice) invariance directly in the architecture. With this approach, the network realizes a systematically improvable superposition of paths of increasing geometric complexity while remaining straightforward to implement. The following subsections introduce the corresponding gauge-equivariant recombinations that serve as our building blocks. Similar layers and networks were developed in Refs.~\cite{Favoni:2020reg,Nagai:2021bhh,Lehner:2023bba}.

The initial assumption is that all open paths of ordered link variables starting and ending at the same point share the same gauge property, and we can define gauge-equivariant recombinations of them. A recurrent combination is the sum of two objects. Let us assume $\phi_1$ and $\phi_2$ transform identically as
\begin{align}
    \phi_1 &\rightarrow G_1\phi_1G_2^\dagger \\
    \phi_2 &\rightarrow G_1\phi_2G_2^\dagger
\end{align}
with some gauge transformations $G_1$ and $G_2$ with $G_i\in\mathrm{SU}(3)$ and $\phi_i\in\mathbb{C}^{3\times 3}$, the set of complex $3\times3$ matrices.
The sum of both objects preserves the gauge transformation
\begin{align}
    \phi_s = \phi_1 + \phi_2 &\rightarrow G_1\phi_1G_2^\dagger + G_1\phi_2G_2^\dagger \\&= G_1(\phi_1+\phi_2)G_2^\dagger = G_1\phi_sG_2^\dagger\label{eq:sum_gauge_equivariant}
\end{align}
where the final object is a superposition of different paths with different shapes or any more complicated object, while the start and end points are preserved. Gauge-equivariant layers rely on this property, with each term in the sum assigned a unique weight that represents the layer's parameters. These weights are independent of the field configuration, specific lattice site, and the orientation of the Wilson loop. This independence ensures the gauge, translational, and rotational invariance of our setup.

The focus in this study lies on on-axis separations between $\bv{x}$ and $\bv{y}$, i.e., $\bv{y}=\bv{x}+r\hat{\mathbf{e}}_r$, where $\hat{\mathbf{e}}_r$ is the direction of separation and $(\bv{x},t)$ a generic lattice site.
However, the procedure can be generalized to off-axis separations as well, which we postpone to future studies.

In the case of on-axis separations, the simplest gauge-variant object is the straight Wilson line connecting $\bv{x}$ and $\bv{y}$, which we denote as $\phizero(x)$; in the notation of Eq.~\eqref{eq:spatial_wilson_line_forward}, $\phizero(x)=S(\bv{x},\bv{x}+\bv{r},t)$.
The next level of straightforward gauge-equivariant modifications involves inserting plaquettes at one of the lattice sites along the straight Wilson line.
There are 24 possible oriented plaquettes in the spatial subvolume containing each site.

A modified Wilson line is given by the expression
\begin{align}
    \phi_{l,i,j}(x) = &\left[\Pi_{k=0}^{l}U_r(\bv{x}+k\hat{\mathbf{e}}_r;t)\right] U_{i,j}(\bv{x}+l\hat{\mathbf{e}}_r;t)\nonumber\\
        &\times \left[ \Pi_{k=l}^{r-1}U_r(\bv{x}+k\hat{\mathbf{e}}_r;t) \right]
    \label{eq:plaquette_insertion_definition}
\end{align}
where $U_{i,j}(x)=U_{i,j}(\mathbf{x};t)$ is an oriented plaquette in the $i$-$j$ plane at site $x$ defined as
\begin{align}
    U_{i,j}(x) = U_i(x)U_j(x+\hat{e}_i)U_{-i}(x+\hat{e}_i+\hat{e}_j)U_{-j}(x+\hat{e}_j)
\end{align}
and the indices run as $l=0,1,\ldots,r$ and $i,j\in \{1,2,3,-1,-2,-3\}$. At insertion locations $l\neq 0$, the four plaquette insertions with a link variable pointed in $-\hat{\mathbf{e}}_r$ direction are already accounted for by the plaquette insertions at $l-1$. Hence, we obtain $24 + 16r$ independent plaquette insertions for a given distance $r$.

A multi-index $i$, running from 1 to $24+16r$, labels all plaquette insertions and is denoted as $\phizero_i(x)$. To maintain 
consistency, we need to establish a convention for the order of insertions. This becomes crucial if the direction of separation 
changes even though the physics remains invariant.
Therefore, we define the order of coordinates as $(1,2,3)$ for $\hat{\mathbf{e}}_r=\hat{\mathbf{e}}_1$, and we adjust the order of 
the coordinates for different directions $(\mu_1,\mu_2,\mu_3)$ such that the $\epsilon$~tensor is 1, i.e., 
$\epsilon_{\mu_1\mu_2\mu_3}=1$. Consequently, for $\hat{\mathbf{e}}_r=\hat{\mathbf{e}}_2$ the order becomes $(2,3,1)$, and for 
$\hat{\mathbf{e}}_r=\hat{\mathbf{e}}_3$ it is $(3,1,2)$.
This convention ensures a consistent ordering of $\phi_{i}^{(0)}(x)$, keeping the neural-network parameters unchanged when evaluating for a different direction. The plaquette-inserted Wilson lines serve as inputs to the neural network (indicated by the $(0)$ superscript), and their recombination through the equivariant layers returns a more complex expression.

For the rest of this work, the subscript of $\phin_i$ ($\phinpo_i$) is an integer with $i=1,\ldots,N\nsub$($i=1,\ldots,N\nposub$) 
where $N\nsub$ ($N\nposub$) is also sometimes referred to as the number of channels.
It is fixed only for $n=0$ ($N\osub=24+16r$ plaquette insertions), while $n>0$ generally indicates the $n$th layer. $N\nsub$ may vary between layers, but must be 1 for the final layer if we consider only a single final state.

We now introduce several layers that are used in the neural networks designed below.
A \emph{linear layer} is a sum of input elements defined as
\begin{align}\label{eq:linear_layer}
    \phinpo_i(x) &= \sum_{j} w\nposub_{ij}\phin_j(x) - b\nposub_{i}\phizero(x),
\end{align}
with the weights $w\nposub\in\mathbb{K}^{N\nposub\times N\nsub}$ and $b\nposub\in\mathbb{K}^{N\nposub}$ for $\mathbb{K}=\mathbb{R}$ or $\mathbb{K}=\mathbb{C}$.
The term with $b_i$ serves as the bias term.
In comparison to a regular linear layer, the bias term needs to be multiplied by the straight Wilson line to maintain gauge equivariance of the layer; this aspect allows us to identify it as the neutral element within our algebra. With Eq.~\eqref{eq:sum_gauge_equivariant} and assuming the input elements obey the required gauge property, this layer is gauge equivariant.

The derivative with respect to the weights of the layers exists and is analytic; hence, automatic differentiation computes it during the backward pass. However, if $\mathbb{K}=\mathbb{C}$, the derivatives with respect to the real and imaginary parts of the weights have to be performed independently.

The purpose of the linear layer is to produce superpositions of plaquette insertions within the Wilson lines, which are associated with cloverlike insertions of the field-strength components $F_{\mu\nu}$.
Field-strength insertions create excited states characterized by a unique quantum number.
Consequently, this layer enhances the expressiveness of excited states, which we aim to minimize in the final output.

Another motivation for this layer is to reduce the number of channels if we do not expect such a high number of degrees of freedom. For example, the initial $24+16r$ possible plaquette insertions consume a high amount of memory, but the actual objects are the field insertions. Hence, an initial linear layer may be used to reduce the number of channels before proceeding with the next layer.

A \emph{bilinear layer} combines the input elements bilinearly; however, only the insertion of the straight Wilson line within the bilinear combination maintains gauge equivariance.
Therefore, we define the bilinear layer as
\begin{align}\label{eq:bilinear_layer}
    \phinpo_i(x) = &\sum_{j,k} w\nposub_{ijk} \phin_j(x)(\phizero(x))^\dagger\phin_k(x)\nonumber\\
    &+ \sum_j \tilde{w}\nposub_{ij}\phin_j(x) - b\nposub_i\phizero(x)
\end{align}
with the weights $w\nposub\in\mathbb{K}^{N\nposub\times N\nsub\times N\nsub}$, $\tilde{w}\nposub\in\mathbb{K}^{N\nposub\times N\nsub}$, and $b\nposub\in\mathbb{K}^{N\nposub}$ for $\mathbb{K}=\mathbb{R}$ or $\mathbb{K}=\mathbb{C}$. The second and third terms correspond to the linear layer and are optional. However, they preserve the contributions of the input elements, which is the case if the bilinear objects cover only partially the underlying problem. The overall motivation for this layer is, for example, that the output elements are proportional to multiple plaquette insertions when the initial elements consist of single plaquette insertions.
Thus, this layer introduces structure along the separation axis.

The bilinear layer is gauge equivariant, as can be demonstrated by a straightforward calculation. Furthermore, $\phinpo_i$ is analytically differentiable with respect to its weights, and, hence, it is calculated in the backward process.
However, the bilinear layer has the disadvantage of producing numerically unstable objects.
The objects $\phin_i$ are not necessarily $\text{SU}(3)$ matrices and might have very small or large numbers, which become even smaller or larger after applying the bilinear layer repeatedly.
This may quickly exceed the float number range, leading to instability during optimization.
Therefore, the use of the bilinear layer requires some care.
Avoiding bilinear layers and using exponential activation functions might be a more stable and self-regulated approach.

A \emph{convolutional layer} combines objects from neighboring lattice sites using parallel transport to maintain gauge equivariance.
Hence, the convolutional layer is defined as
\begin{widetext}
\begin{align}
    \phinpo_i(x) &= \sum_{j} \sum_{\mu=0,\pm 1,\pm 2, \pm 3} w\nposub_{ij\mu} U_{\mu}(x) \phin_j(x+\hat{e}_\mu) U_{-\mu}(x+\hat{e}_\mu+\mathbf{r}) - b\nposub_{i}\phizero(x)\label{eq:convolutional_layer_analytical_definition}\\
    &= \sum_j\Bigg\{\sum_{\mu=1,2,3}\Big[\tilde{w}_{ij\mu}\nposub U_{\mu}(x) \phin_j(x+\hat{e}_\mu) U^\dagger_{\mu}(x+\mathbf{r})+\hat{w}_{ij\mu}\nposub U^\dagger_{\mu}(x-\hat{e}_\mu) \phin_j(x-\hat{\mu}) U_{\mu}(x+\mathbf{r}-\hat{\mu})\Big] \nonumber\\
    &\ \hspace{1.5cm} + w_{ij,0}\nposub \phin_j(x) \Bigg\} - b\nposub_{i}\phizero(x) \label{eq:convolutional_layer_technical_definition}
\end{align}
\end{widetext}
with $\tilde{w}\nposub_{ij\mu}=w\nposub_{ij\mu}$ and $\hat{w}\nposub_{ij\mu}=w\nposub_{ij,-\mu}$, and $w\nposub\in\mathbb{K}^{N\nposub\times N\nsub\times 7}$, $b\nposub\in\mathbb{K}^{N\nposub}$ for $\mathbb{K}=\mathbb{R}$ or $\mathbb{K}=\mathbb{C}$. 
Equation~\eqref{eq:convolutional_layer_analytical_definition} defines the convolutional layer, while 
Eq.~\eqref{eq:convolutional_layer_technical_definition} is a small change in notation for the weights corresponding to the technical implementation. The last term corresponds to the bias term and is optional.
Note that in Eq.~\eqref{eq:convolutional_layer_analytical_definition} the case $\mu=0$ represents a shorthand notation for no convolution, which is explicitly denoted in the second line of Eq.~\eqref{eq:convolutional_layer_technical_definition}.
Thus, the term containing $w\nposub_{ij,0}$ generates a linear layer that preserves structures obtained in the previous layer.

The convolutional layer introduces structure orthogonal to the separation axis.
Although it does not produce explicitly objects proportional to rectangular plaquette insertions, it produces 
objects with a prolonged extension perpendicular to the separation axis, and the output object is a superposition that might have rectangular plaquette insertions.
The exact contribution of the individual objects is, however, governed by the tunable weights. 
Furthermore, the output objects can be interpreted as fat links obtained similarly in smearing procedures such as APE 
smearing~\cite{APE:1987ehd}.
The key difference is that the exact shape of the smearing is parametrized by the weights, while in APE smearing a specific choice is made by hand.
As with the linear and bilinear layers, the convolutional layer is a parametrized gauge-equivariant recombination and is analytically differentiable with respect to the weights.

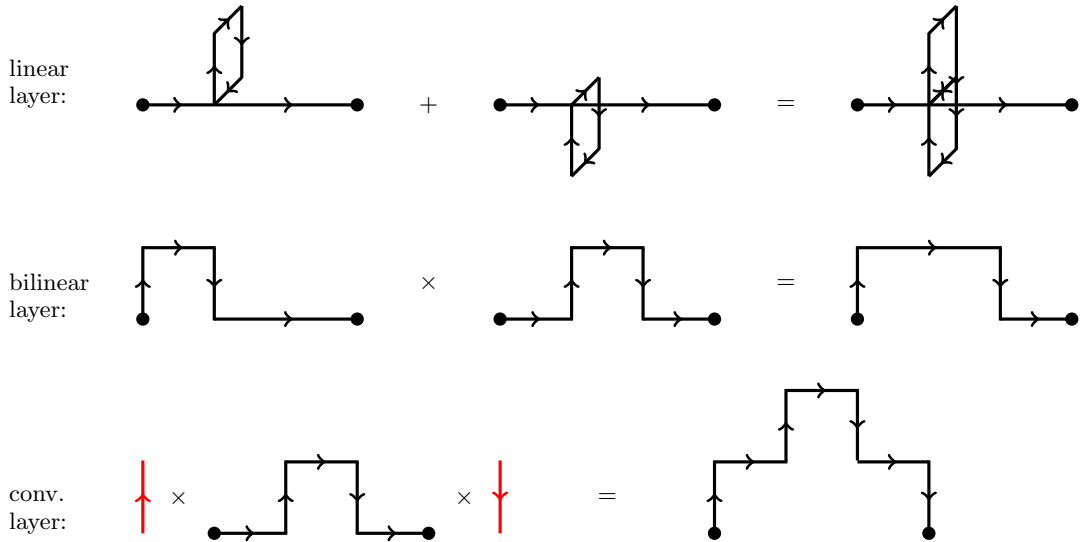
\begin{figure*}
    \centering
    \begin{tikzpicture}[scale=0.95]


\begin{scope}[very thick,decoration={
    markings,
    mark=at position 0.55 with {\arrow{>}}}
    ] 
    \draw[postaction={decorate}] (0,0)--(1,0);
    \draw[postaction={decorate}] (1,0)--(3,0);
    
    \draw[postaction={decorate}] (1,0)--(1,1.01);
    \draw[postaction={decorate}] (1,1,0)--(1,1,-1);
    \draw[postaction={decorate}] (1,1,-1)--(1,-0.01,-1);
    \draw[postaction={decorate}] (1,0,-1)--(1,0,0);

    \filldraw[black] (0,0) circle (2pt) node[anchor=west]{};
    \filldraw[black] (3,0) circle (2pt) node[anchor=west]{};
    
\end{scope}

\node[text width=4.5cm] at (0.5,0.3){linear\\ layer:};

\node[] at (4,0) {$+$};

\begin{scope}[very thick,decoration={
    markings,
    mark=at position 0.55 with {\arrow{>}}}
    ] 
    \draw[postaction={decorate}] (5,0)--(6,0);
    \draw[postaction={decorate}] (6,0)--(8,0);

    \draw[postaction={decorate}] (6,0)--(6,0,-1);
    \draw[postaction={decorate}] (6,0,-1)--(6,-1.01,-1);
    \draw[postaction={decorate}] (6,-1,-1)--(6,-1,0);
    \draw[postaction={decorate}] (6,-1,0)--(6,0,0);

    \filldraw[black] (5,0) circle (2pt) node[anchor=west]{};
    \filldraw[black] (8,0) circle (2pt) node[anchor=west]{};
    
\end{scope}

\node[] at (9,0) {$=$};

\begin{scope}[very thick,decoration={
    markings,
    mark=at position 0.55 with {\arrow{>}}}
    ] 
    \draw[postaction={decorate}] (10,0)--(11,0);
    \draw[postaction={decorate}] (11,0)--(13,0);

    \draw[postaction={decorate}] (11,0)--(11,1.01);
    \draw[postaction={decorate}] (11,1,0)--(11,1,-1);
    \draw[postaction={decorate}] (11,1,-1)--(11,-1.010,-1);
    \draw[postaction={decorate}] (11,0,-1)--(11,0,0);
    
    \draw[postaction={decorate}] (11,0)--(11,0,-1);
    \draw[postaction={decorate}] (11,0,-1)--(11,-1.01,-1);
    \draw[postaction={decorate}] (11,-1,-1)--(11,-1,0);
    \draw[postaction={decorate}] (11,-1,0)--(11,0,0);

    \filldraw[black] (10,0,0) circle (2pt) node[anchor=west]{};
    \filldraw[black] (13,0,0) circle (2pt) node[anchor=west]{};

\end{scope}


\begin{scope}[very thick,decoration={
    markings,
    mark=at position 0.55 with {\arrow{>}}}
    ] 
    \draw[postaction={decorate}] (0,-3)--(0,-1.98);
    \draw[postaction={decorate}] (0,-2)--(1.02,-2);
    \draw[postaction={decorate}] (1,-2)--(1,-3.02);
    \draw[postaction={decorate}] (1,-3)--(3,-3);
    \filldraw[black] (0,-3) circle (2pt) node[anchor=west]{};
    \filldraw[black] (3,-3) circle (2pt) node[anchor=west]{};    
\end{scope}

\node[text width=4.5cm] at (0.5,-2.7){bilinear\\ layer:};

\node[] at (4,-2.5) {$\times$}; 

\begin{scope}[very thick,decoration={
    markings,
    mark=at position 0.55 with {\arrow{>}}}
    ] 
    \draw[postaction={decorate}] (5,-3)--(6.02,-3);
    \draw[postaction={decorate}] (6,-3)--(6,-1.98);
    \draw[postaction={decorate}] (6,-2)--(7.02,-2);
    \draw[postaction={decorate}] (7,-2)--(7,-3.02);
    \draw[postaction={decorate}] (7,-3)--(8,-3);
    \filldraw[black] (5,-3) circle (2pt) node[anchor=west]{};
    \filldraw[black] (8,-3) circle (2pt) node[anchor=west]{};
\end{scope}

\node[] at (9,-2.5) {$=$};

\begin{scope}[very thick,decoration={
    markings,
    mark=at position 0.55 with {\arrow{>}}}
    ] 
    \draw[postaction={decorate}] (10,-3)--(10,-1.98);
    \draw[postaction={decorate}] (10,-2)--(12.02,-2);
    \draw[postaction={decorate}] (12,-2)--(12,-3.02);
    \draw[postaction={decorate}] (12,-3)--(13,-3);
    \filldraw[black] (10,-3) circle (2pt) node[anchor=west]{};
    \filldraw[black] (13,-3) circle (2pt) node[anchor=west]{};
\end{scope}


\begin{scope}[very thick,decoration={
    markings,
    mark=at position 0.55 with {\arrow{>}}}
    ] 
    \draw[color=red,postaction={decorate}] (0,-6)--(0,-4.98);
    \draw[color=red,postaction={decorate}] (5,-4.98)--(5,-6);
    \draw[postaction={decorate}] (1,-6)--(2.02,-6);
    \draw[postaction={decorate}] (2,-6)--(2,-4.98);
    \draw[postaction={decorate}] (2,-5)--(3.02,-5);
    \draw[postaction={decorate}] (3,-5)--(3,-6.02);
    \draw[postaction={decorate}] (3,-6)--(4,-6);
    \filldraw[black] (1,-6) circle (2pt) node[anchor=west]{};   
    \filldraw[black] (4,-6) circle (2pt) node[anchor=west]{};
\end{scope}

\node[] at (0.5,-5.5) {$\times$};
\node[] at (4.5,-5.5) {$\times$};

\node[text width=4.5cm] at (0.5,-5.7){conv.\\ layer:};

\node[] at (6.5,-5.5) {=};

\begin{scope}[very thick,decoration={
    markings,
    mark=at position 0.55 with {\arrow{>}}}
    ] 
    \draw[postaction={decorate}] (8,-6)--(8,-4.98);
    \draw[postaction={decorate}] (8,-5)--(9.02,-5);
    \draw[postaction={decorate}] (9,-5)--(9,-3.98);
    \draw[postaction={decorate}] (9,-4)--(10.02,-4);
    \draw[postaction={decorate}] (10,-4)--(10,-4.98);
    \draw[postaction={decorate}] (10,-5)--(11,-5);
    \draw[postaction={decorate}] (11,-4.98)--(11,-6);
    \filldraw[black] (8,-6) circle (2pt) node[anchor=west]{};
    \filldraw[black] (11,-6) circle (2pt) node[anchor=west]{};
\end{scope}

\end{tikzpicture}
    \caption{Sketch of the actions of linear (top), bilinear (middle), and convolutional (bottom) layers, showing how gauge equivariance is built in.
    The black dots represent static quark-antiquark pairs, while the arrowed lines show the path connecting them.
    The linear layer creates clover-leaf-like insertions; the bilinear layer generates structure along the separation axis; and the convolutional layer creates structure orthogonal to the separation axis.}
    \label{fig:layer_sketch}
\end{figure*}
Figure~\ref{fig:layer_sketch} illustrates examples of how the linear, bilinear, and convolutional layers create gauge-equivariant structure through iterative application.

An \emph{$\text{SU}(3)$ exponential activation} is a parameter-free layer mapping any complex $3\times 3$ matrix onto SU(3). The technical implementation consists of two steps, which are outlined below. The first step projects out the traceless, anti-Hermitian part of $\phin_j(\phizero(x))^\dagger$, an object that transforms locally at~$x$, as
\begin{align}
    \tilde{\phi}^{(n)}_i(x) &= \phin_i(x)(\phizero(x))^\dagger - \left(\phin_i(x)(\phizero(x))^\dagger\right)^\dagger \nonumber\\
    &\  -\frac{1}{3}\tr [\phin_i(x)(\phizero(x))^\dagger - (\phin_i(x)(\phizero(x))^\dagger)^\dagger].
\end{align}
$\tilde{\phi}\nsub$ inherits the gauge property of $\phin_j(\phizero(x))^\dagger$, and it is an element of $\mathfrak{su}(3)$.
Hence, an exponentiation of this object returns an $\text{SU}(3)$ matrix.
The second step multiplies the exponential with $\phizero$,
\begin{align}
    \phinpo_i(x) &= e^{\tilde{\phi}^{(n)}_i(x)}\phizero(x).
\end{align}
The exponential of a locally gauge-transforming object remains a locally gauge-transforming object, which can be verified by representing the exponential as a power series. The final multiplication with $\phizero$ preserves the gauge equivariance.

The SU(3) exponential activation function enjoys two advantages. The first one includes combinations of the objects 
similar to bilinear, trilinear, and $n$-linear operations with decreasing weights while still producing $\text{SU}(3)$ matrices 
that are regulated. That means they have well-behaved numbers that do not encounter floating-point issues while capable of similar 
complexity. The second is based on the regularization property; any final state of the neural network has a clear norm, which we  
discuss below when constructing the loss function.

The matrix exponential is an analytic operation, so it is differentiable with respect to the dependent variables.
We use the Fr\'echet derivative~\cite{al2009computing,bader2019computing} implemented in the built-in routines of PyTorch, which computes the derivative of a scalar and real loss function for the steepest descent with respect to real-valued parameters. This qualifies it for gradient descent methods, which optimize for a defined loss function.

In summary, the linear, bilinear, convolutional, and exponential activation layers provide a set of gauge-equivariant building blocks of a feedforward neural network.
Starting with simple plaquette insertions that obey the correct gauge property, the output of the neural network 
exhibits the same gauge property and can be understood as a sum of differently shaped paths. The contributions of the individual paths are parametrized by the layer's weights and can be tuned such that the neural network optimizes a chosen objective function. 
The level of complexity is tuned by the number of hidden layers, the number of hidden channels, and the choice of the layers. 
These choices constitute a class of hyperparameters.

\subsection{Neural-network architecture}
\label{sec:network_architecture}

Section~\ref{sec:layers_general} defines gauge-equivariant layers which have the same gauge property as the straight Wilson 
line defined in Eq.~\eqref{eq:spatial_wilson_line_forward}.
Here we use the layers to parametrize a static quark-antiquark state, 
replacing the spatial Wilson lines in Eq.~\eqref{eq:Wilson_loop_definition}. The spatial part corresponds to a single superposition 
of multiple paths connecting $\bv{x}$ to $\bv{x}+\bv{r}$; therefore, the number of channels of the final layer is equal to 1 (i.e., $N^{(n_\text{final})}=1$) in the regular case. We might denote it with
\begin{align}
    \tilde{S}(\bv{x},t)=\phi_1^{(n_\text{final})}(x)
\end{align}
and in Eq.~\eqref{eq:Wilson_loop_definition} we replace both $S(x)$, at time slices~$0$ and~$t$, with $\tilde{S}(x)$, keeping the straight temporal Wilson lines.

We write the traced neural-network parametrized generalization of the Wilson loop as $\langle \tr \widetilde{W}_{r\times t}(\{w^{(n)}\})\rangle$, explicitly 
including the dependence on the weights, because they serve as the parameters of our variational \emph{Ansatz}.
For the rest of the study, the expression $\langle \tr \widetilde{W}_{r\times t}\rangle$ contains the dependence on the weights implicitly. The expectation value of the Wilson loop is real valued, but it might have a remnant imaginary part of the order of the floating-point precision.
Therefore, it is safe to take the real part in numerical work.

The norm of a state is defined as $\langle \tilde{S}|\tilde{S}\rangle\equiv \langle \tr \widetilde{W}_{r\times 0}\rangle=\sum_x\langle \tr \tilde{S}(x)\tilde{S}^\dagger(x)\rangle/V$ with the lattice volume $V$. If $\tilde{S}(x)$ is an SU(3) matrix, the norm will always be equal to 3. In the case of a parametrized superposition of different paths, the explicit norm depends on the weights; however, the norm does not affect the spectral quantity of the observable, and it plays only a role in the numerical stability of the calculation.

To build a network, we can choose either a \emph{flat} network with only a few hidden layers or a \emph{deep} network with many hidden layers. It turns out that a flat network converges faster to a saturation point in the loss function, while a deep network converges slowly or not at all. In addition, many initial bilinear layers may encounter numerical problems with numerical zeros or not-a-numbers (NaNs) because they are related to high-degree polynomials of the lattice field numbers.
On the other hand, a flat network might not capture the full complexity of gluon dynamics.
The dynamics have an effective range corresponding to the radius of the flux tube; each application of a convolutional layer increases the effective range of the network by one lattice spacing. Therefore, we need to find a balance between flat and deep network architectures.
We employ progressive learning; namely, we begin with a flat network and add new layers iteratively in a controlled manner.

The general network construction procedure is as follows: In the first step, we set up an initial network. The first layer consists of a linear layer with $N^{(1)} = 16$ or $N^{(1)} = 14$.\footnote{In the course of this work, we have not carried out a detailed study of the neural networks' dependence on $N^{(1)}$.}
Since we do not expect all initial plaquette insertions to be independent, this layer effectively reduces the number of elements, which saves memory on the hardware required for the backward call.
After the initial layer, two to four different combinations of convolutional, bilinear, and exponential layers follow, where we keep the number of input and output elements constant and equal for the hidden layers. We initialized the weights randomly, such that the initial network remains close to the straight Wilson line.

In the second step, we optimize the network with respect to a loss function for a specified number of epochs, reaching a saturation point in the loss function. We add a new layer one before the final layer in the third step, initiated with random weights. To adjust the new weights, we freeze the previous weights and perform a few epochs only with the new weights before continuing with the full training. We repeat steps two and three until we reach a limit. These limits may be due to memory bounds or a loss of training stability. By storing all intermediate networks, we can select the optimal network for our purpose by inspecting the loss function as it evolves over the course of the number of epochs. This simple procedure works only if the number of input and output elements of the hidden layers is the same.

\subsection{Loss function}

To optimize the network towards a physically meaningful result, we need to motivate loss functions that aim for the physics we are interested in. In quantum many-body physics, a variational \emph{Ansatz} $|\Psi(\{ w^{(n)}\}\rangle$ parametrized by a set of weights $\{w^{(n)}\}$ is optimized to approach the ground state of the system by minimizing the energy expectation value, i.e.,
\begin{align}
    |\Psi_0\rangle &\approx |\Psi(\{w^{(n)}\}_0)\rangle , \nonumber \\
    \{w^{(n)}\}_0 &= \argmin_{\{w^{(n)}\}}  \frac{\langle\Psi(\{w^{(n)}\})|H|\Psi(\{w^{(n)}\})\rangle}{\langle\Psi(\{w^{(n)}\})|\Psi(\{w^{(n)}\}\rangle}. \label{eq:ground_state_hamilton_definition}
\end{align}
Neural network quantum states (NQS) are a class of variational \emph{Ansätze} to approximate the ground state of a system by using a strategy to minimize Eq.~\eqref{eq:ground_state_hamilton_definition}; see, for example~\cite{Carleo:2016svm}. However, an NQS provides only an approximation of the ground state, and further inspections of its correctness are needed. In our study, we aim to approximate the ground state of the static quark-antiquark system. In contrast to NQS, an approximation is sufficient, as the Euclidean time propagation converges to the ground state. In this case, a good ground-state approximation leads to stabler fitting at earlier Euclidean-time separations, which is crucial for precision calculations.

In lattice-gauge-theory computations, the energy expectation value $\langle\Psi(\{w\})|H|\Psi(\{w\})\rangle$ is not directly accessible. Instead, a lattice correlator provides the matrix elements of the transfer matrix in Eq.~\eqref{eq:transer_matrix}, with $T=e^{-aH}$.
Since the transfer matrix is an exponential of the negative Hamiltonian, and this function is strictly monotonically decreasing, Eq.~\eqref{eq:ground_state_hamilton_definition} can be reformulated to 
\begin{align}
    |\tilde{S}_0\rangle &\approx |\tilde{S}(\{w^{(n)}\}_0)\rangle , \nonumber \\
    \{w^{(n)}\}_0 &= \argmax_{\{w^{(n)}\}} \frac{\langle \tilde{S}(\{w^{(n)}\},t)|\tilde{S}(\{w^{(n)}\},0)\rangle}{\langle \tilde{S}(\{w^{(n)}\})|\tilde{S}(\{w^{(n)}\})\rangle}.
    \label{eq:ground_state_transfer_matrix}
\end{align}
We identify $\tilde{S}(\{w^{(n)}\},t)$ as our neural-network Wilson line parametrized by the set $\{w^{(n)}\}$ at time slice $t$ and $\langle\tr \widetilde{W}_{r\times t}\rangle\equiv \langle \tilde{S}(\{w^{(n)}\},t)|\tilde{S}(\{w^{(n)}\},0)\rangle$.
It follows that $\langle\tr \widetilde{W}_{r\times 0}\rangle\equiv \langle \tilde{S}(\{w^{(n)}\})|\tilde{S}(\{w^{(n)}\})\rangle$.

Equation~\eqref{eq:ground_state_transfer_matrix} thus provides an objective function for Euclidean correlators which, when the parameters are optimized with respect to it, delivers an improvable approximation of the initial ground state.
Two ambiguities remain, however, in the final formulation of the loss function.
One is conceptual, since we can state independent conditions for each $t$, while the second is numerical, since $\langle \tilde{S}(\{w^{(n)}\})|\tilde{S}(\{w^{(n)}\})\rangle$ is not uniquely fixed.

We split the loss function for $\langle\tr \widetilde{W}_{r\times t}\rangle$ in a physical loss function $L^\mathrm{phys}$ which encodes Eq.~\eqref{eq:ground_state_transfer_matrix} and its ambiguity in $t$, and a regulator loss function $L^\mathrm{reg}$ which regulates the ambiguity of $\langle \tilde{S}(\{w^{(n)}\})|\tilde{S}(\{w^{(n)}\})\rangle$ and provides numerical stability while having no impact on the physical outcome. The final loss function is given by
\begin{align}
    L=L^\mathrm{phys} + L^\mathrm{reg}\label{eq:loss_function_general_definition}
\end{align}
with the individual terms being given by
\begin{align}
    L^\mathrm{phys}&=-\sum_{t=1}^{t_\mathrm{max}}W_t\frac{\langle\tr \widetilde{W}_{r\times t}\rangle}{\langle \tr \widetilde{W}_{r\times 0}\rangle},
    \label{eq:loss_function_phys} \\
    L^\mathrm{reg}&=W\left(\langle\tr \widetilde{W}_{r\times 0}\rangle - N\right)^2 ,
    \label{eq:loss_function_reg} \\
    \sum_{t=1}^{t_\mathrm{max}}W_t&= t_\mathrm{max},
    \label{eq:Wt_normalization_convention}
\end{align}
where $t_\mathrm{max}$, $W_t$, $N$, and $W$ are hyperparameters, to be discussed further below. The minus sign in $L^\mathrm{phys}$ recasts the maximizing condition, Eq.~\eqref{eq:ground_state_transfer_matrix}, as a minimizing problem, as is standard in machine learning.
The hyperparameters $t_\mathrm{max}$, $W_t$, and $N$ play the following roles: $t_\mathrm{max}$ sets the maximum time difference, once the signal at larger times becomes statistically weak; $W_t$ is used to fine-tune the optimization window, because excited-state contamination is more dominant at small $t$, while noise dominates at larger $t$; $N$ sets the target norm; and $W$ adjusts the importance of the norm condition. Equation~\eqref{eq:Wt_normalization_convention} is a simple normalization convention based on the case where $W_t=1$, a uniformly weighted window.
Given this formulation of the loss function, we see that $L^\mathrm{phys}$ drives the network toward its ground state, while $L^\mathrm{reg}$ prevents the network from running into numerical instabilities emerging in numerical zeros or NaNs.

The derivation of the loss function can also be obtained directly from the spectral representation in Eq.~\eqref{eq:spectral_representation_Wilson_loop}.
If $\tilde{c}_n$ [for $c_n\rightarrow\tilde{c}_n$ in Eq.~\eqref{eq:spectral_representation_Wilson_loop}, while the $E_n$ remain the same]
are the overlap coefficients of the neural-network state with the $n$th state, the optimal case is $\tilde{c}_0\approx1$ and $\tilde{c}_{n>0}\approx0$. However, $\langle\tr \widetilde{W}_{r\times t}\rangle$ does not provide direct access to $\tilde{c}_n$, so the loss function cannot directly optimize for the ideal case $\tilde{c}_0^\mathrm{ideal}=1$, $\tilde{c}_{n>0}^\mathrm{ideal}=0$.
Instead, it is necessary to argue through the correlator itself, as we do in the following.

We start by inspecting the ideal case, which is only the ground state without excited-state contamination, given by a single exponential decay as
\begin{align}
    \frac{\langle\tr \widetilde{W}_{r\times t}^\mathrm{ideal}\rangle}{\langle\tr \widetilde{W}_{r\times 0}^\mathrm{ideal}\rangle} = e^{-taE_0(r)}
\end{align}
where, in general, $\tilde{c}_0^\mathrm{ideal}\neq 1$ but we focus on $\tilde{c}_0^\mathrm{ideal}=1$, which is enforced by the regulator term, and $\tilde{c}_{n>0}^\mathrm{ideal}=0$. The realistic case, however, is given by
\begin{align}
        \frac{\langle\tr \widetilde{W}_{r\times t}\rangle}{\langle\tr \widetilde{W}_{r\times 0}\rangle} = \frac{\sum_{n=0}^\infty |\tilde{c}_n|^2e^{-taE_n(r)}}{\sum_{n=0}^\infty |\tilde{c}_n|^2}.
\end{align}
The ideal case defines our optimum; therefore, the loss function must minimize the difference between the ideal case and the actual case.

The difference between the ideal and the actual case is
\begin{align}
    \Delta_t \equiv e^{-taE_0(r)} - \frac{\sum_{n=0}^\infty |\tilde{c}_n|^2e^{-taE_n(r)}}{\sum_{n=0}^\infty |\tilde{c}_n|^2}.
\end{align}
From this, we can make some simple assumptions:
\begin{align}
    |\tilde{c}_n| &\geq 0\\
    E_{n>0}(r) &\geq E_{0}(r) > 0
\end{align}
from which it follows $0<e^{-ta(E_{n>0}(r)-E_0(r))}<1$ and thus
\begin{align}
    \Delta_t &= e^{-taE_0(r)}\left( 1- \frac{\sum_{n=0}^\infty |\tilde{c}_n|^2e^{-ta(E_{n>0}(r)-E_0(r))}}{\sum_{n=0}^\infty |\tilde{c}_n|^2}\right) 
    \nonumber \\
    &> 0.
\end{align}
Because the difference $\Delta_t$ is positive for all $t$, the ratio $\langle\tr \widetilde{W}_{r\times t}\rangle/\langle\tr \widetilde{W}_{r\times 0}\rangle$ is smaller than the ideal case.
Therefore, maximizing this ratio corresponds to minimizing $\Delta_t$, eventually converging to a correlator dominated by the ground state. Additionally, summing over all $t$ with hyperparameters~$W_t$ restores $L^\mathrm{phys}$. To recover the full loss function, we introduce $L^\mathrm{reg}$ for the same reason as in the transfer matrix derivation. We finally obtain the same loss function from two different starting points of our argumentation: one from the Hamiltonian and transfer matrix description, and one from the spectral representation of the correlator.

Both chains of arguments leading to the loss function assume a positive transfer matrix.
To preserve positivity, we retain the original temporal link variables and remain within the spatial subvolume to prevent any temporal overlap, thereby guaranteeing positivity. If these conditions do not hold, we would have to reformulate the loss function, which is  not considered in this study.\footnote{In the case of improved actions, we tacitly assume that physical positivity~\cite{Luscher:1984is} suffices.}
Since the whole argumentation relies only on the structure of a Euclidean time correlator, its principle can be generalized to other states, such as gluelumps, glueballs, or any hadron. Developing the corresponding neural networks is left for future research.

Before concluding this section, we discuss alternative loss functions that lead to the same physics.
The first alternative is to minimize or flatten the effective mass directly. Minimizing corresponds to minimizing the energy, and flattening corresponds to reducing excited-state contributions. However, to achieve a good estimate for the effective mass for a single gradient descent step, many field configurations need to be evaluated, typically on the order of $100$.
This requires computational time, and the signal is lost at larger separations $r$, which slows down the training process. Equation~\eqref{eq:loss_function_phys} for defining $L^\mathrm{phys}$ also works for much smaller batch sizes; in fact, estimating it with a single lattice configuration for each gradient descent step turns out to be enough for computational cost-efficient training.

Another alternative is to minimize the variance of the Wilson loop directly. A robust determination of the variance requires, however, numerous samples, leading again to a more costly training regimen. Additionally, variance reduction does not guarantee a correlator dominated by the ground state.

\section{Applications and results}
\label{sec:results}

Having introduced the theoretical and technical background in the previous sections, in this section, we proceed to a realistic test problem.
In Sec.~\ref{sec:static_energy}, we introduce, test, and compare different network architectures on two spatial separations of the Wilson loops.
We study which network architecture works well and extend it to train it serially on all separations.
After training, we freeze the weights of the optimized network and measure this observable on the full ensemble, and, finally, extract the static energies.
In Sec.~\ref{sec:static_force}, we will apply the optimized network to measure the static force with an operator consisting of a Wilson loop with a chromoelectric field insertion in one of the temporal Wilson lines, which is a valuable playground to test the networks for additional observables. We follow this with a brief demonstration of combining the neural-network method with the multilevel algorithm in Sec.~\ref{sec:multilevel}, which sets the foundations for further precision calculations.
In Sec.~\ref{sec:excited_states}, we extend the formalism for single ground states to a tower of excited states and let the network find automatically interpolators approximating excited states.
In Sec.~\ref{sec:unveiling_network}, we visualize the learned structure of the network by tracing the operations of the network in an explicit algebraic representation of the paths.

In this study, we train with pure-gauge configurations on a $20^3\times40$ lattice with the Wilson gauge-field action at $\beta=6.281$. 
The training ensemble consists of 6000 configurations used previously~\cite{Brambilla:2023fsi}.
We randomly draw a single configuration set for each optimization iteration. For the final measurement, we generated an additional set of 20,000 lattice configurations, using the overrelaxation and pseudo-heat-bath algorithms in the MILC code~\cite{MILC}.

The training and final measurement program was developed in Python using PyTorch, which uses the automatic differentiation module and the AdamW~\cite{loshchilov2019decoupledweightdecayregularization} optimizer. The backward algorithm for the convolutional and partially the exponential activation is explicitly implemented. This saves memory, as the standard implementation requires multiple slicing calls, which do not align with the underlying structure of the layers. Conceptual tests were conducted on a single NVIDIA RTX 4090 GPU, while serial training and evaluation were performed on a cluster with NVIDIA H100 GPUs.

\subsection{Ground-state optimization}
\label{sec:static_energy}

There are many possibilities for constructing a network architecture. We vary the number of layers, layer types, and the number of hidden channels to investigate how different choices affect performance.
Our approach involves starting with a fixed network topology and then iteratively adding new layers, as outlined in Sec.~\ref{sec:network_architecture}, while varying the number of hidden channels. We will compare the best results obtained from networks with different configurations that have a comparable number of hidden channels. This will allow us to select the optimal network based on both training quality and computational cost.

In the first iteration, we study a network architecture of alternating convolutional and bilinear layers.
The initial network consists of four layers: linear $\rightarrow$ convolutional $\rightarrow$ bilinear $\rightarrow$ convolutional.
(Henceforth, we abbreviated convolutional with ``conv'' and bilinear with ``bilin.'')
New ``ConvBilin'' pairs of layers are added iteratively; the choice of a different number of hidden channels and the corresponding label is listed in Table~\ref{tab:ConvBilin_alternating}.
\begin{table}
    \centering
    \caption{Labels for the convolutional-bilinear (ConvBilin) alternating networks with the corresponding number of hidden channels. $N^{(0)}$ is 72 for $r=3$ and 136 for $r=7$, and $N^{(n_\text{final})}=1$. The initial network architecture is linear $\rightarrow$ conv $\rightarrow$ bilin $\rightarrow$ conv.}
    \label{tab:ConvBilin_alternating}
    \begin{tabular}{ccc}
       \hline\hline
       Label  & $N^{(1)}$ & $N^{(n>1)}$ \\
       \hline
       ConvBilin\_1  & 14 & 4\\
       ConvBilin\_2  & 14 & 6\\
       ConvBilin\_3  & 14 & 8\\
       ConvBilin\_4  & 14 & 10\\
       \hline\hline
    \end{tabular}
\end{table}

\begin{figure}
    \centering
    \includegraphics[width=0.45\textwidth]{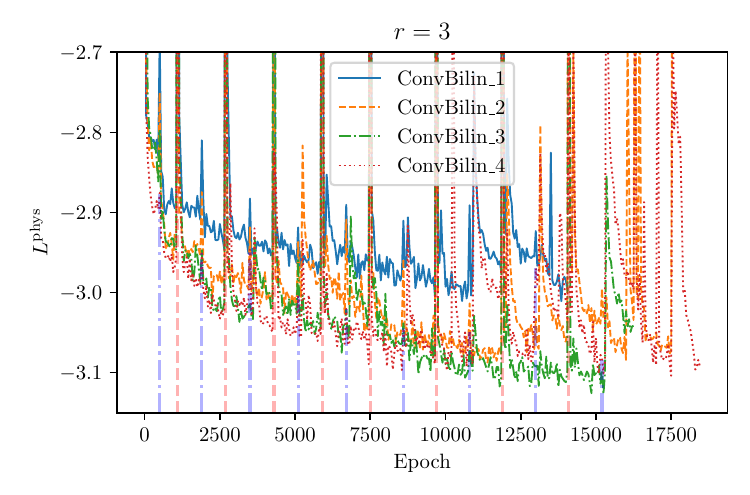}
    \includegraphics[width=0.45\textwidth]{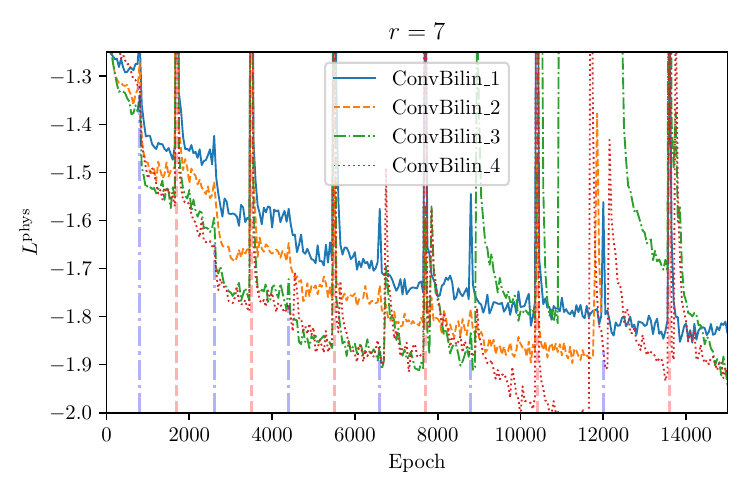}
    \caption{Comparing the training history for the ConvBilin training procedure. Blue vertical lines indicate the insertion of a convolutional layer, while the red lines indicate the insertion of a bilinear layer. The $x$ axis represents the epoch, i.e., the training time, while the $y$ axis displays the value for $L^\mathrm{phys}$ only (not the full loss function).
    A~lower value for $L^\mathrm{phys}$ corresponds to a more optimized result.}
    \label{fig:ConvBilin_training_history}
\end{figure}

Figure~\ref{fig:ConvBilin_training_history} shows the training history of $L^\mathrm{phys}$ for $r=3$ and $r=7$. We observe a discontinuous change (spike) every time a new ConvBilin pair is added. This behavior is to be expected, as adding a new layer introduces new parameters that push the value of the loss function far away from its current optimal value.
Additional training steps are obviously required to thermalize the new setup.

We further see that increasing the number of hidden channels increases the performance of the network; the impact at $r=7$ is larger as at $r=3$, indicating that for larger distances a higher number of parameters supports the quality of the signal.
Too many parameters, here for $N^{(n>1)}$, lead, however, to instability in the training, which is, again, more likely at $r=7$ than at $r=3$. This phenomenon occurs due to increased statistical fluctuations at greater distances, while managing more parameters becomes more challenging. Additionally, we find that higher network complexity does not yield significantly better results beyond a certain point. Apart from the weight decay in the AdamW algorithm, no further regularization procedures, such as clipping of the gradient norm, have been applied, an aspect that should be studied in future research.

\begin{figure}
    \centering
    \includegraphics[width=0.45\textwidth]{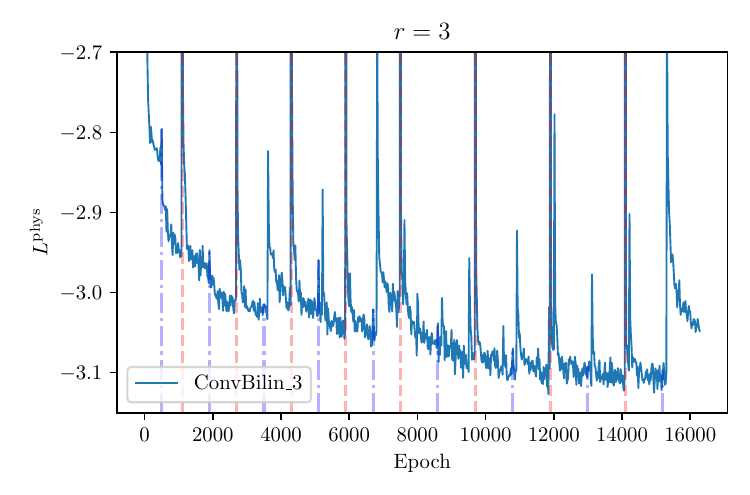}
    \includegraphics[width=0.45\textwidth]{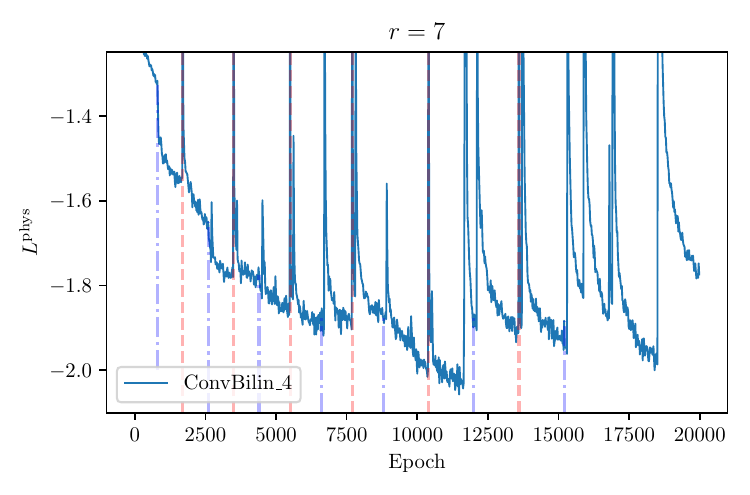}
    \caption{The training histories for each $r$ of a single network. Blue vertical lines indicate the insertion of a convolutional layer, while the red lines indicate the insertion of a bilinear layer. The $x$ axis represents the epoch, i.e., the training time, while the $y$ axis displays the value for $L^\mathrm{phys}$ only (not the full loss function).
    A~lower value for $L^\mathrm{phys}$ corresponds to a more optimized result.}
    \label{fig:Single_ConvBilin_training_history}
\end{figure}

Figure~\ref{fig:Single_ConvBilin_training_history} shows the training history of a single network for better visibility. We find that both types of insertions, convolutional and bilinear layers, provide a significant boost in training performance and finally reach a saturation point. Furthermore, we cannot observe a qualitative difference between the insertion of a convolutional or bilinear layer. However, in our experience, the bilinear layer causes numerical instabilities if not treated with great care. Therefore, we will focus on networks without bilinear layers for the rest of this study.

We introduce two additional network concepts: the ``ConvExp'' and the ``ConvLimNeighbor.''  The ConvExp network consists solely of convolutional layers, aside from the initial linear layer, with each layer followed by an exponential activation function. This design leverages the regularization properties and multilinear structure of the exponential action.

In contrast, the ConvLimNeighbor network adopts a different philosophy. Instead of adding new convolutional layers with a fixed number of hidden channels, it incorporates a convolutional layer with fewer hidden channels that operates solely over adjacent sites. The new objects are integrated with the preceding layer by combining it with a linear layer that restores the standard number of hidden channels. The initial weights of this linear layer are chosen such that the new construction acts like an identity. This approach also allows the new weights to be randomly initialized, while the weights of the additional linear layer are set to ensure that the new convolutional layer has a neutral effect. This avoids the potential issue of a symmetric gradient for the new weights, which, although rare, could be problematic. 

Moreover, the network's capacity can be expanded by adding more neighboring convolutional layers that require fewer weights than those typically expected with the standard number of hidden layers. We anticipate greater statistical fluctuations as the network becomes thicker, and we do not expect to achieve high information density from distant convoluted lattice sites. This design provides additional stability during training for the ConvLimNeighbor network architecture.

\begin{figure}
    \centering
    \includegraphics[width=0.45\textwidth]{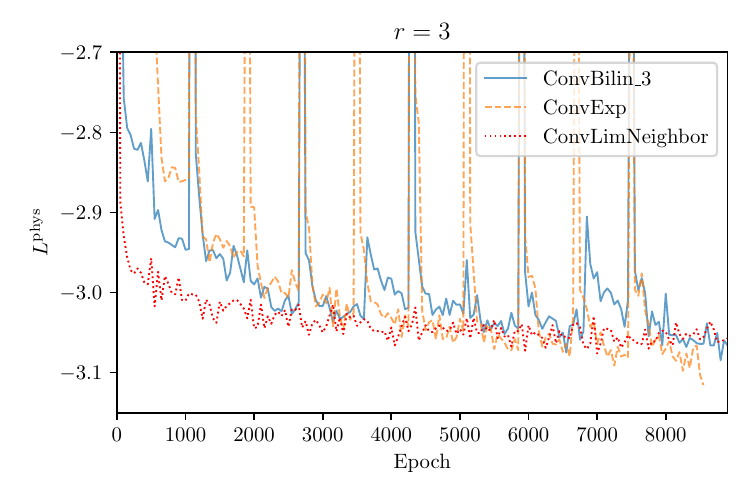}
    \includegraphics[width=0.45\textwidth]{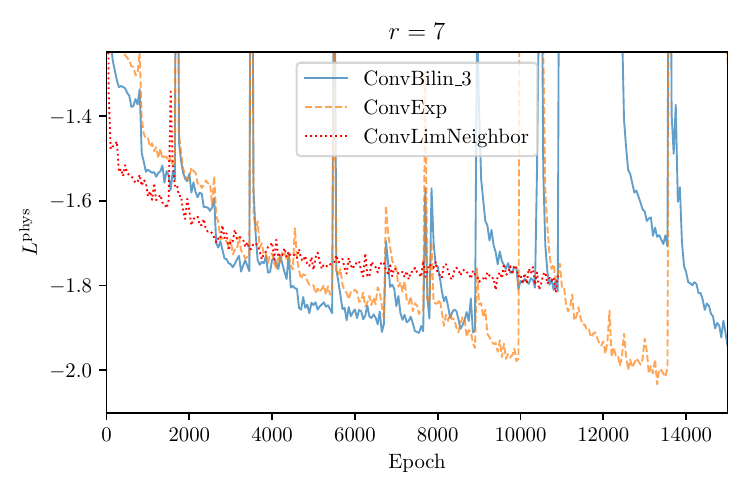}
    \caption{The training histories for each $r$ of three different network architecture concepts. We obtain similar peak structures for the ConvExp as for the previous ConvBilin network. In contrast, the ConvLimNeighbor exhibits stability over a wide range in the training history.}
    \label{fig:Single_more_nets_training_history}
\end{figure}

Figure~\ref{fig:Single_more_nets_training_history} shows the training history of the new network concepts and compares them with the previous ConvBilin networks. We observe that the ConvExp network behaves similarly to the ConvBilin networks; however, it performs better at $r = 7$, although it remains challenging to control. In contrast, the ConvLimNeighbor network demonstrates reliable stability over a wide range of training steps. Nevertheless, for $r = 7$, the ConvLimNeighbor network does not achieve optimal performance in the loss function compared to the other networks. Nonetheless, stability is a crucial property for serially training networks across various $r$ values without the need for fine-tuning each individual case.

To cover all values of $r$, we choose the ConvLimNeighbor network architecture and train a separate network for each value of $r$ ranging from 1 to 10. As this network architecture is stable, we increase the number of hidden channels as $r$ increases, assuming that this leads to better performance for larger values of $r$. We add new layers iteratively until the training becomes unstable or the program runs out of memory, at which point it fails automatically. Each network is saved before a new layer is added, allowing us to select a set of networks for final evaluation and comparison. For this computation, we freeze the weights and treat the network as an observable, measuring it across the full ensemble.

\begin{figure}
    \centering
    \includegraphics[width=0.45\textwidth]{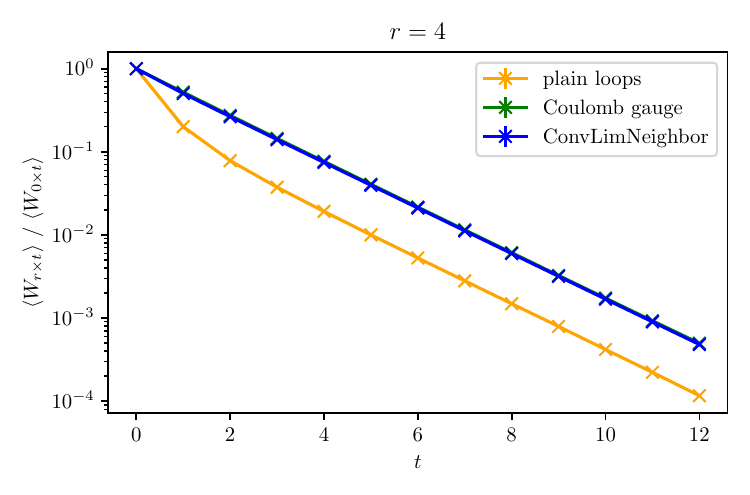}
    \includegraphics[width=0.45\textwidth]{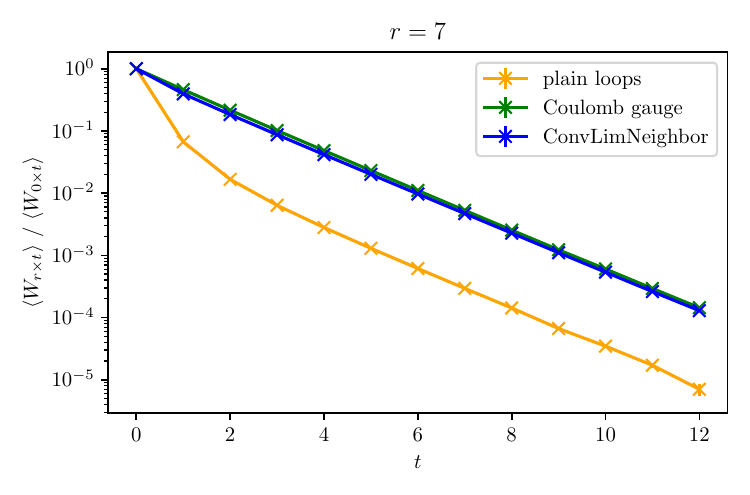}
    \caption{The bare, normalized correlator from the final measurement for $r=4$ and $r=7$ in a logarithmic plot. The data points are connected with lines to guide the eye.}
    \label{fig:loop_correlator_comparison}
\end{figure}

Figure~\ref{fig:loop_correlator_comparison} presents the bare, normalized correlator from the final measurement for $r = 4$ and $r = 7$. We compare the results obtained from the neural network with those from Coulomb-gauge-fixed Wilson-line correlators and plain Wilson loops. While the plain Wilson loops are affected by excited-state contamination at small $t$, the network correlators and Coulomb-gauge Wilson lines appear nearly identical in a logarithmic plot, suggesting a dominant single exponential decay behavior. For larger $t$, all three correlators exhibit the same decay rate, indicating that they represent the same ground state. Notably, especially in the $r = 7$ plot, the signal for the plain Wilson loops deteriorates more quickly than that of the neural-network Wilson loops and Coulomb-gauge Wilson lines. In summary, our observations suggest that both the neural network and the Coulomb-gauge Wilson-line correlator yield comparable results.

\begin{figure}
    \centering
    \includegraphics[width=0.45\textwidth]{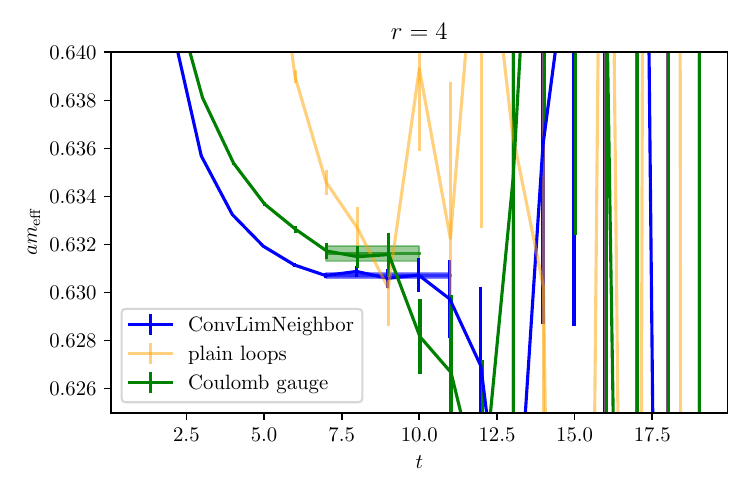}
    \includegraphics[width=0.45\textwidth]{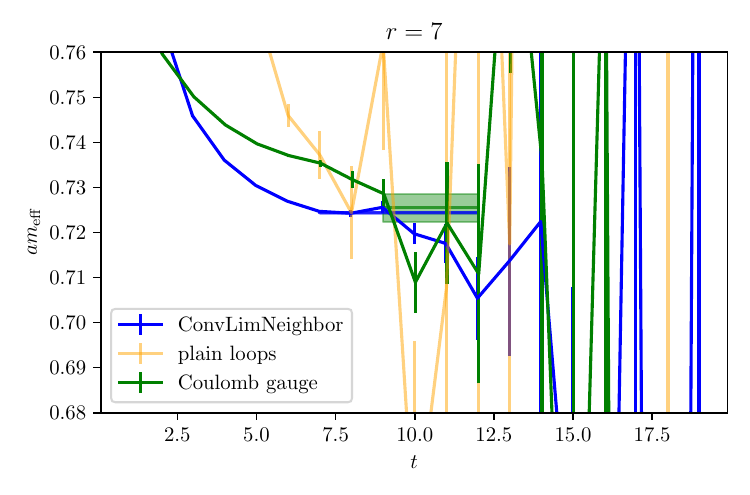}
    \caption{The effective masses for the different correlators at $r=4$ and $r=7$.}
    \label{fig:meff_comparison}
\end{figure}

Next, we analyze the effective masses of the correlators defined by the following equation:
\begin{align}
    am_\mathrm{eff}(t) = -\ln \frac{\langle \tr W_{r \times (t+1)} \rangle}{\langle \tr W_{r \times t} \rangle} ,
\end{align}
where $W$ represents either the neural-network Wilson loop, the plain Wilson loop, or the Coulomb-gauge Wilson lines.
Figure~\ref{fig:meff_comparison} illustrates the effective mass plots for the three correlators. Both the neural-network Wilson loop and the Coulomb-gauge Wilson lines demonstrate a significantly better signal and much less excited-state contamination than the plain Wilson loops. However, at small $t$ (up to five or six), the Coulomb-gauge Wilson lines and the neural network still suffer from some excited-state contamination. Furthermore, the neural network performs comparably well or even better than the Coulomb-gauge Wilson lines in terms of signal-to-noise ratio and the range of the plateau. Therefore, we conclude that the neural-network approach is as effective as the Coulomb-gauge method while still providing gauge-invariant operators. Reducing further the excited-state contamination at small $t$ remains an open challenge for future studies.

\begin{figure}
    \centering
    \includegraphics[width=0.45\textwidth]{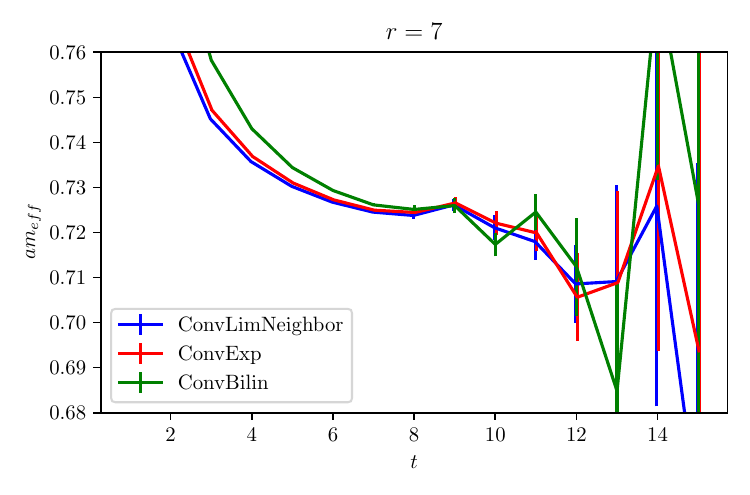}
    \caption{The effective mass for different network architectures.}
    \label{fig:meff_comparison_2}
\end{figure}

A relevant question is how different network architectures influence their final physical outcomes. So far, we have introduced three different network architectures. Figure~\ref{fig:meff_comparison_2} shows the effective mass for all three. We notice that all three networks show the same plateau range and value. The primary distinction lies in the excited-state contamination at small $t$, which we aim to minimize. The difference between the ConvBilin network and the other two architectures disappears when a more complex ConvBilin network is used; this configuration was selected for demonstration purposes. Since the final physical outcome depends only slightly on the network architecture, the optimal network is assessed based on signal quality and computational cost. More complex networks typically require greater computational resources. However, in certain situations, prioritizing networks that minimize excited-state contamination---regardless of computational cost---can be advantageous.

To extract the static energy of the ground state
\begin{align}
aE_0(r) = \lim_{t\rightarrow\infty} am_\mathrm{eff}(t)|_r,
\end{align}
we perform a plateau fit by selecting a reasonable time range and propagating the statistical error through blocking and jackknife resampling. An example of this analysis is shown in Fig.~\ref{fig:meff_comparison}, showing statistical uncertainties only.

\begin{figure}
    \centering
    \includegraphics[width=0.45\textwidth]{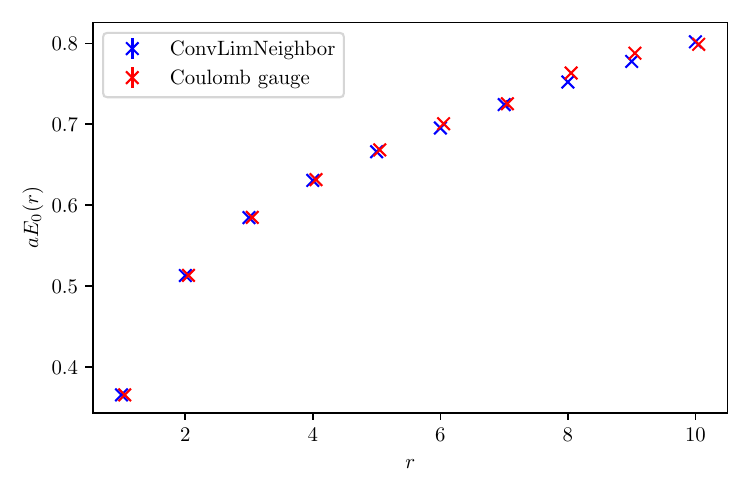}
    \caption{The static energy as a function of the distance $r$ for the neural-network approach in comparison to the Coulomb-gauge approach. Statistical error bars are included, but not visible.}
    \label{fig:static_energy_groundstate}
\end{figure}

Figure~\ref{fig:static_energy_groundstate} presents a comparison of the static energy obtained from the neural-network approach and the Coulomb-gauge method. Both methods yield similar results, although discrepancies arise at larger values of $r$.
These discrepancies can be attributed to the typical systematic differences from excited-state contamination that arise when comparing different correlators.

We conclude that the neural-network method introduced in this study is as effective as the Coulomb-gauge method. This is significant because the neural network provides a gauge-invariant operator that can be used for further processing and can be applied in situations where implementing the Coulomb gauge is not straightforward. Furthermore, the neural-network approach can be integrated with other signal-improving techniques, such as hypercubic (HYP) smearing~\cite{Hasenfratz:2001hp}, gradient flow, and multilevel algorithms. A different application of the optimized operator is to apply it directly to ground-state expectation values, or as an input state for a correlation matrix measurement. We will explore those possibilities in the following sections.

The static energies extracted from the plateau fit define the ideal value of the loss function. So far, we have optimized the loss function without knowing its theoretically optimal value. This situation corresponds to a single exponential of the correlator, where the mass precisely matches the static energy. Therefore, by inserting this exponential function with the static energy in the exponent into the loss function, we represent the ideal case, which we then compare to the value obtained from the measured correlator.

Table~\ref{tab:Lphysmeas_optimum_comparison} compares the measured loss function $L^\mathrm{MPhys}$ with the optimal value $L^\mathrm{OPhys}$. We take differences $\Delta L = L^\mathrm{OPhys} - L^\mathrm{MPhys}$ and ratios $L^\mathrm{OPhys}/L^\mathrm{MPhys}$ of both values. The loss function serves as an absolute measure, where a value of 0 indicates a correlator without any correlation. Its minimum value is defined by the static energy exponential. Thus, the ratio of the measured to the ideal value reflects the relative success of the neural network.
\begin{table}
    \centering
    \caption{Comparison of the final, measured physical loss function with the theoretical optimal value obtained from the mean values of the effective masses.}
    \label{tab:Lphysmeas_optimum_comparison}
    \begin{tabular}{ccccc}
        \hline\hline
        $r$	& $-L^\mathrm{MPhys}$ & $-L^\mathrm{OPhys}$ & $-\Delta L$ & $\frac{L^\mathrm{MPhys}}{L^\mathrm{OPhys}}$ \\
        \hline
        1	& 5.171 &  5.205 &  0.034 &  0.993 \\
        2	& 3.636 &  3.712 &  0.076 &  0.979 \\
        3	& 3.081 &  3.210 &  0.130 &  0.960 \\
        4	& 2.758 &  2.939 &  0.181 &  0.939 \\
        5	& 2.478 &  2.751 &  0.273 &  0.901 \\
        6	& 2.193 &  2.611 &  0.419 &  0.840 \\
        7	& 1.974 &  2.481 &  0.507 &  0.796 \\
        8	& 1.715 &  2.363 &  0.648 &  0.726 \\
        9	& 1.438 &  2.266 &  0.828 &  0.634 \\
        10	& 1.147 &  2.177 &  1.030 &  0.527 \\
        \hline\hline
    \end{tabular}
\end{table}

We observe that for $r=1$ the success rate is nearly 1, gradually decreasing to around 0.5 as $r$ increases. This trend suggests that more complex networks may be necessary for larger distances, as one can expect more intricate gluon dynamics in such cases. Further improvements to the networks will be explored in future studies.

\subsection{Static force as a ground-state expectation value}
\label{sec:static_force}

The static force is defined as the derivative of the static energy:
\begin{align}
    F_{\partial_r}^\mathrm{cont}(r) &= \frac{\partial E_0(r)}{\partial r} .
\end{align}
This quantity is physical and therefore ultraviolet and infrared finite.
On the lattice, the static force is approximated using a numerical derivative of the static energy. For example, by applying a symmetric two-point derivative, we can express it on the lattice as
\begin{align}
    a^2 F_{\partial_r}(r) &= \frac{1}{2}\left[aE_0(r+1) - aE_0(r-1)\right] .
    \label{eq:static_foce_derivative}
\end{align}
In this equation, $aE_0(r)$ represents the static energy measured from the $t \rightarrow \infty$ limit of the effective masses of the Wilson loops.

The static force is used for scale setting through the indirect equation
\begin{align}
    r^2F_{\partial_r}(r) |_{r=r(c)} &= c.
\end{align}
A common choice is the Sommer scale~\cite{Sommer:1993ce} defined as $r_0\equiv r(1.65)$, which is widely used to set the lattice spacing. Other common choices are $r_1\equiv r(1)$~\cite{Bernard:2000gd} and $r_c\equiv r(0.65)$.

This definition is accompanied by systematic errors stemming from the numerical derivative of the static energy. Therefore, an alternative definition was introduced in~\cite{Vairo:2015vgb,Vairo:2016pxb,Brambilla:2000gk} as
\begin{align}
        F_E(r)&=-\lim_{t\rightarrow\infty}i\frac{\langle \tr \{ W_{r\times t}\hat{\bv{r}\cdot}g\bv{E}(\bv{r},t^\ast)\}\rangle}{\langle \tr W_{r\times t} \rangle}\nonumber\\
    &= -\lim_{t\rightarrow\infty}i\frac{\langle\tr W\!E_{r\times t}\rangle}{\langle\tr W_{r\times t}\rangle}
    \label{eq:static_force_E}
\end{align}
where $\hat{\bv{r}}$ is the direction vector of the quark-antiquark separation, $\bv{E}$ is the chromoelectric field vector, $t^*$ is an arbitrary time slice, and $g$ is the bare coupling.
The object in the numerator, with the shorthand notation $W\!E_{r\times t}$, is a Wilson loop of spatial extent $\bv{r}$ with a chromoelectric field insertion at $t^\ast$ in the Wilson line at $\bv{r}$.
The notation implies, but does not spell out, the path ordering of the modified Wilson loop.

The $t\rightarrow\infty$ limit yields ground-state dominance, so $F_E(r)$ converges to the ground-state expectation value of the chromoelectric field~\cite{Brambilla:2021wqs}.
Therefore, this object can be interpreted as
\begin{align}
    F_E(r) &= \frac{\langle\Psi_0|-i\hat{\bv{r}}\cdot \bv{E}(\bv{r})|\Psi_0\rangle}{\langle\Psi_0|\Psi_0\rangle}
\end{align}
where $|\Psi_0\rangle$ is the ground state of the static quark-antiquark pair separated by $\bv{r}$.
In previous studies, e.g., Refs.~\cite{Brambilla:2021wqs,Brambilla:2023fsi}, only the plain Wilson loop was considered as the trial state, but we expect a better overlap could arise from optimizing a generalized Wilson loop.
Thus, we investigate here how we can use the neural-network methods developed above to improve the overlap with the ground state, thereby enhancing the overall measurement.

On the lattice, we discretize the $\bv{E}$~field, and multiple choices exist. We pick the clover leaf discretization; for details, see Ref.~\cite{Brambilla:2023fsi}.
These discretizations require renormalization, introduced as
\begin{align}
    Z_EF_E(r) = F_{\partial r}(r) .
    \label{eq:E_field_renormalization}
\end{align}
In principle, $Z_E$ is independent of $r$, while practical implementations tend to show a mild, residual dependence.
A nonperturbative definition of $Z_E$ is thus
\begin{align}
    Z_E = \frac{F_{\partial r}}{F_E} .
\end{align}
In this study, we will focus only on the comparison between measuring the static force with and without the neural-network-optimized Wilson loops.
A previous study used the gradient flow method to renormalize the $\bv{E}$~field~\cite{Brambilla:2023fsi}.

\begin{figure}
    \centering
    \includegraphics[width=0.45\textwidth]{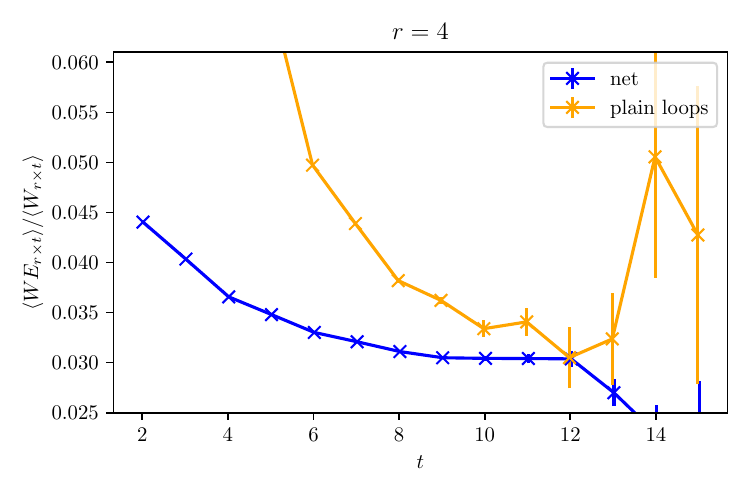}
    \includegraphics[width=0.45\textwidth]{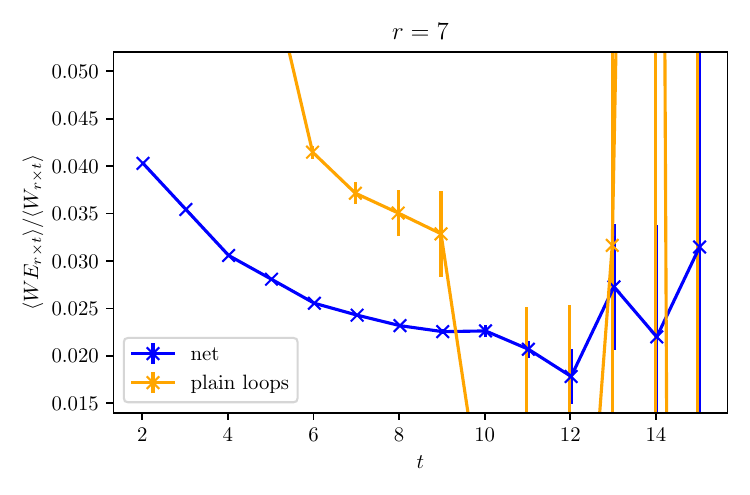}
    \caption{The $t$ dependence of the $\bv{E}$-field insertion in the Wilson loop for two different distances $r$. The figures show the comparison of the ConvLimNeighbor-optimized Wilson loop with the plain Wilson loop. We observe that the quantities approach a plateau, where the neural-network Wilson loop exhibits a significantly improved signal and a more robust plateau. The neural network is clearly superior to the plain Wilson loop.}
    \label{fig:E_field_insertions_Wloops}
\end{figure}

We measure the $\bv{E}$-field insertion in the ConvLimNeighbor-optimized Wilson loop and compare it with the $\bv{E}$-field insertion in the plain Wilson loop. Figure~\ref{fig:E_field_insertions_Wloops} shows a comparison between both measurements.
We observe a significant improvement in the signal with the neural-network Wilson loop while both measurements approach the same value.

To extract the $t\rightarrow\infty$ limit, we carry out a plateau fit akin to the effective mass extraction used for determining the static energy. We select an appropriate range in $t$ and perform constant fits for the neural-network Wilson loops. However, the signal from the plain Wilson loops does not provide sufficient data for a reliable plateau extraction.

\begin{figure}
    \centering
    \includegraphics[width=0.45\textwidth]{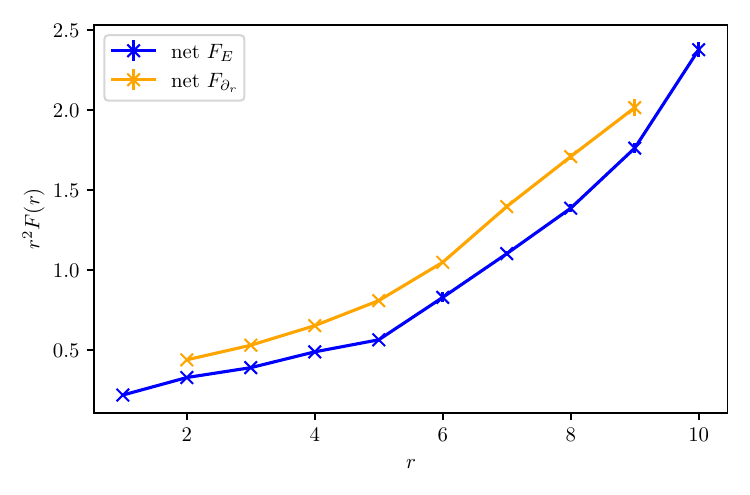}
    \caption{Comparison of the static force obtained through the numerical derivative, Eq.~\eqref{eq:static_foce_derivative}, and the chromoelectric-field insertion, Eq.~\eqref{eq:static_force_E}. We notice a multiplicative constant shift between both results, which is attributed to the renormalization $Z_E$ of the $\bv{E}$~field introduced in Eq.~\eqref{eq:E_field_renormalization}.}
    \label{fig:static_force_final_comparison}
\end{figure}

Figure~\ref{fig:static_force_final_comparison} shows the static force as a function of $r$, obtained via both Eqs.~\eqref{eq:static_foce_derivative} and~\eqref{eq:static_force_E}. In both cases, the neural-network Wilson loop is shown. We observe a multiplicative constant shift between both curves, which is expected due to the required renormalization $Z_E$ of the $\bv{E}$~field introduced in Eq.~\eqref{eq:E_field_renormalization}.

We have demonstrated that neural-network Wilson loops can significantly enhance the measurement of ground-state expectation values for the chromoelectric field.
Because the improvement stems from the faster approach to the ground state, we expect the enhancement will hold for various other operators too.
Additionally, this method is especially beneficial in situations---as here with the static-force matrix element---where applying the Coulomb gauge is not straightforward.
Overall, improving the measurement of matrix elements is crucial for various applications in lattice QCD, spanning from static force calculations to hadronic expectation values.

\subsection{Neural networks with the multilevel algorithm}
\label{sec:multilevel}

In this section, we provide a brief overview of how the neural-network method can be integrated with existing algorithms and methods. In particular, we demonstrate how the multilevel algorithm~\cite{Luscher:2001up} can be used to further improve the signal. The multilevel algorithm applies here because we only study the pure gauge theory using the ultralocal Wilson action. The multilevel algorithm can be applied to the temporal links because the neural network affects only the spatial part.

After optimizing the Wilson loop with a neural network, the largest source of statistical fluctuation is found to originate from the temporal Wilson lines. In the multilevel algorithm, we sample the temporal-link variables $N_\mathrm{ml}$ additional times. Since the Wilson action is local, we can treat each time slice independently from the other time slices. The final Wilson loop is constructed by combining all possible temporal slices, which increases the effective sample size by a factor of order $N_\mathrm{ml}^t$ for Wilson loops of temporal extent~$t$.

The multilevel object is computed from multilevel temporal link variables
\begin{align}
    \mathbb{U}_{ABCD}(\bv{x},\bv{r},t) \equiv \hspace{14em} \nonumber \\ \hspace{4em}
    \frac{1}{N_\mathrm{ml}}\sum_{i=1}^{N_\mathrm{ml}} U_4^{(i)}(\bv{x}+\bv{r},t)_{AB}\ U_4^{(i)\dagger}(\bv{x},t)_{CD} ,
\end{align}
where $i$ sums over the different additionally sampled time slices, and the indices $A$, $B$, $C$, and $D$ are the indices for the components of the $3\times 3$ matrices. The multilevel temporal-link variables form the multilevel temporal Wilson line as
\begin{align}
    \mathbb{T}_{ABCD}(\bv{x},\bv{r},t) &= \mathbb{T}_{AEFB}(\bv{x},\bv{r},t-1)\mathbb{U}_{ECDF}(\bv{x},\bv{r},t) ,
\end{align}
where the Einstein summation rule applies, and at the start
\begin{align}
    \mathbb{T}_{ABCD}(\bv{x},\bv{r},t)|_{t=1} &= \mathbb{U}_{ABCD}(\bv{x},\bv{r},0) .
\end{align}
The multilevel-improved Wilson line is computed as
\begin{align}
    \tr \widetilde{W}_{r\times t}|_\mathrm{ml} = &\tilde{S}(\bv{x},\bv{x}+\bv{r},0)_{AB}\nonumber\\
    &\times\mathbb{T}_{BCDA}(\bv{x},\bv{r},t)\tilde{S}^\dagger(\bv{x},\bv{x}+\bv{r},t)_{CD}
\end{align}
where $\tilde{S}$ can also be replace by the straight line $S$.

We use the 6000 lattice configuration employed for training as initial configurations to sample additional $N_\mathrm{ml}=20$ temporal-link variables. Since the spatial part with the neural network is composed only of spatial link variables, we need to compute it only for the initial configurations and keep it fixed for the contraction of the multilevel temporal Wilson lines. Thus, this method is an efficient approach to improve the signal.

\begin{figure}
    \centering
    \includegraphics[width=0.45\textwidth]{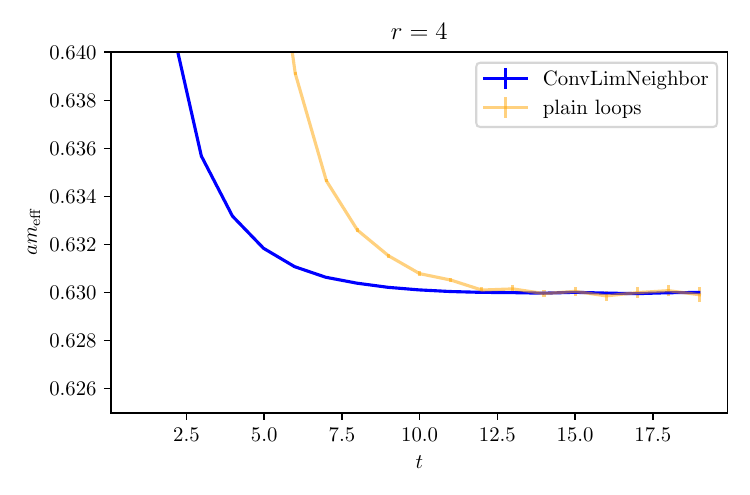}
    \includegraphics[width=0.45\textwidth]{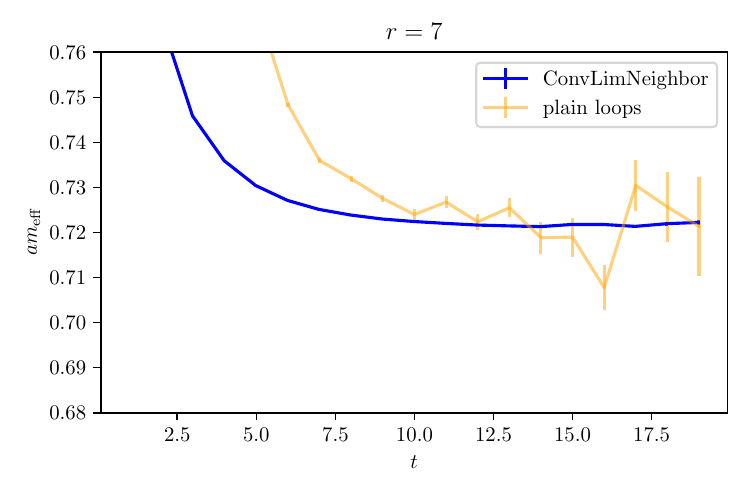}
    \caption{Two examples of the effective masses for the multilevel-improved Wilson loops. It shows the multilevel-improved Wilson loops for the plain Wilson loop and the neural-network Wilson loop.}
    \label{fig:meff_ml_comparison}
\end{figure}

Figure~\ref{fig:meff_ml_comparison} shows two examples of effective masses for the multilevel-improved Wilson loops. We observe a significant improvement for both the plain Wilson loops and the neural-network Wilson loops. At larger distances, here $r=7$, the neural-network Wilson loop preserves a stable signal while the plain Wilson loop is already losing the signal. This example further demonstrates the utility of neural networks for Wilson loops.

\subsection{Seeking  excited states}
\label{sec:excited_states}

In the previous sections, we studied the ground state of the static quark-antiquark pair and optimized the interpolator to minimize excited-state contamination.
The excited states themselves are particularly important for understanding gluon dynamics and the spectrum of exotic states, and they are referred to as hybrid static energies because they model hadrons with a valence gluon and a heavy $Q\bar{Q}$ pair.
They are characterized by a specific set of quantum numbers, related to the symmetry of the two static sources at a fixed distance in QCD.
To obtain hybrid energy levels, the static energies are input for the Born-Oppenheimer effective field theory for hybrids~\cite{Berwein:2015vca,Berwein:2024ztx}. They have been calculated on the lattice since the seminal work of Juge \emph{et al.}~\cite{Juge:1997nc,Juge:1999ie,Juge:2002br}.

Before showing results from our neural networks, let us review the symmetry properties of the spectrum.
On the lattice, these symmetries include four discrete rotations around the separation axis, which correspond to the cubic rotation group: $R_{\pi/2}$, $R_\pi$, $R_{3\pi/2}$, and $R_{2\pi} = \mathds{1}$. Additionally, there are reflections at a plane spanned by the separation axis and an orthogonal direction (for example, $\mathcal{P}_x$ or $\mathcal{P}_y$ if $z$ is the separation axis). There is also a combined operation of charge conjugation followed by a reflection at the center point of the separation axis, ($\mathcal{P} \circ \mathcal{C}$).

The cubic rotations generate the cubic rotation group, whose irreducible representations are related to continuum angular momentum. For the three lowest angular momentum states, the labels $\Lambda = \Sigma$, $\Pi$, $\Delta$ suffice,
where $\Sigma$ corresponds to the $J=0$ state, $\Pi$ to the $J=1$ state, and $\Delta$ to the $J=2$ state. 

The eigenvalues of the reflections $\mathcal{P}_{x/y}$ are labeled by $\epsilon = \pm$. The eigenvalues of the combined operation $\mathcal{P} \circ \mathcal{C}$ are denoted as $\eta = g$ (for even), and $u$ (for odd).
Thus, the complete notation for a state is given by $\Lambda_\eta^\epsilon$.
For example, the static quark-antiquark ground state is denoted $\Sigma_g^+$.

It is important to note that for $J \geq 1$, the two states with $\epsilon=\pm$ are degenerate.
Thus, the states $\Pi_\eta$ and $\Delta_\eta$ form doublets.
To obtain the excited states, interpolators of superpositions of differently shaped paths, obeying the correct quantum numbers, can be constructed; recent extensive studies can be found in~\cite{Capitani:2018rox,Schlosser:2021wnr}.

\begin{figure}
    \centering
    \includegraphics[width=0.45\textwidth]{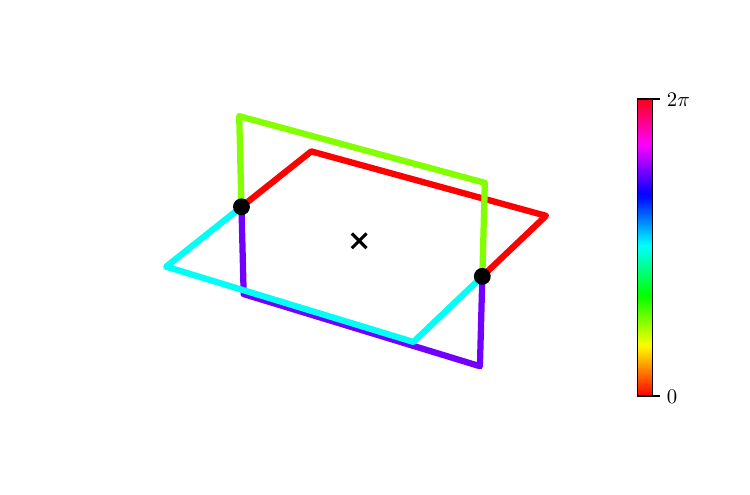}
    \caption{Illustrative picture for a $\Pi_-$ state. The black dots note the static quarks, and the cross marks the center point of the separation. The paths are color coded according to their complex phase.}
    \label{fig:Pi_u_picture}
\end{figure}

The $\Pi _u$ state is the first excited state and therefore of particular interest. It has $J=1$ angular momentum, which is generated by the $\pi/2$ rotations, each assigned with a phase shift of $e^{i\pi/2}=i$. Hence, given a trial path $S_\mathrm{trial}$, the $J=1$ state is constructed by $(\mathds{1}+iR_{\pi/2}-R_{\pi}-iR_{3\pi/2})S_\mathrm{trial}$. An example for a rectangular trial path is shown in Fig.~\ref{fig:Pi_u_picture}. Furthermore, a reflection at the center point yields an overall minus sign, defining the odd property under the $\mathcal{P}\circ\mathcal{C}$ transformation for the $\Pi_u$ state. This pictorial understanding is crucial to understanding Sec.~\ref{sec:unveiling_network}, where we unveil the neural network to see what it has learned.

In general, a defining property of excited states is their orthogonality. Given a tower of states of a theory $|\psi_m\rangle$ numerated by $m$ ($m$ can also be a multi-index, like for the hybrid states), the states fulfill $\langle\psi_{m'}|\psi_m\rangle\propto\delta_{m'm}$. The states are ordered such that $E_{m'}\geq E_m$ if $m'>m$, where $m=0$ defines the ground state and the remaining states are the excited states.
If $E_{m'}=E_m$ even if $m'\neq m$, the states for $m'$ and $m$ are degenerate, as in the doublets for $\Pi_\eta$ and $\Delta_\eta$ discussed above.

The orthogonality leads to two major conclusions. First, the spectral representation Eq.~\eqref{eq:spectral_representation_Wilson_loop} only sums over states with the same quantum number of the initial state, i.e., $c_n=0$ if $n\neq m$. Second, the final, optimized state from Eq.~\eqref{eq:ground_state_transfer_matrix} has the same quantum number as the trial states $\tilde{S}$. In conclusion, we can apply Eqs.~\eqref{eq:loss_function_phys} and~\eqref{eq:loss_function_reg} on a neural network that represents a state with the desired quantum number. Therefore, we can use the established procedure to optimize for the ``ground state'' within the domain of a fixed quantum number.

To obtain neural-network trial states of excited states, we have two options. For the first option, we construct neural networks that explicitly obey a specific quantum number. This method establishes constraints on the network weights and, hence, requires more technical developments. In the second option, we create a set of trial states without any constraints, orthogonalize the set, and apply the loss function for each orthogonalized state individually. This method requires only a minor technical modification, but we cannot determine the explicit quantum numbers without additional measurements. In this work, we will focus on the second option, whereas the first holds promising prospects for future studies.

We modify the neural network by changing the number of final output elements, i.e., $N^{(n_\text{final})}>1$.
The states are therefore labeled $\tilde{S}_i(\bv{x},t)=\phi_i^{(n_\text{final})}(x)$, $i=1, \ldots, N^{(n_\text{final})}$.
Here, we focus specifically on $N^{(n_\text{final})}=8$.
The neural-network Wilson loop composed of $\tilde{S}_i(\bv{x},0)$, $\tilde{S}_j(\bv{x},t)^\dagger$, and the two temporal Wilson lines then yields a correlation matrix as $C_{ij}(t)=\langle\tr\widetilde{W}_{ij,r\times t}\rangle$.
We employ the generalized eigenvalue problem (GEVP)
\begin{align}
    C(t)v_n(t,t_0) &= \lambda_n(t,t_0)C(t_0)v_n(t,t_0)\label{eq:GEVP_definition}
\end{align}
to orthogonalize the set of states~\cite{Michael:1982gb,Kronfeld:1989tb,Luscher:1990ck}, where $n$ labels the tower of orthogonal (excited) states, $\lambda_n$ is the generalized eigenvalue, i.e., the correlation function of the corresponding (excited) state with $\lambda_n(t_0,t_0)=1$, $v_n(t,t_0)$ is the generalized eigenvector, and $t_0$ is a reference time.
We focus here on $t_0=0$; in general, the dependence of the final result on $t_0$ needs to be considered.

The new loss function is defined as
\begin{align}
    L &= L^\mathrm{phys,GEVP}+L^\mathrm{reg}+L^\mathrm{ortho}
\end{align}
where the new individual components are given by
\begin{equation}
    L^\mathrm{phys,GEVP} = \sum_i^{n_\text{final}}W_i^s \left( -\sum_{t=1}^{t_\mathrm{max}} W_t\lambda_i(t,t_0) \right) ,
    \label{eq:loss_excited_states_phys_GEVP}
\end{equation}
which is a generalization of Eq.~\eqref{eq:loss_function_phys} to all excited states with an additional set of hyperparameters $W_i^s$, and
\begin{equation}
    L^\mathrm{ortho} = \sum_{i>j} \left( \frac{\langle\tr \widetilde{W}_{ij,r\times 0}\rangle}{\sqrt{|\langle\tr \widetilde{W}_{ii,r\times 0}\rangle||\langle\tr \widetilde{W}_{jj,r\times 0}\rangle|}} \right)^2 ,
    \label{eq:loss_ortho}
\end{equation}
which enforces orthogonality of the neural network.
The term $L^\mathrm{ortho}$ is not necessary for the physics, but can help to maximize the learned information in the initial set of states.

We employ the previously introduced ConvLimNeighbor architecture, as it has demonstrated high stability in the training history. The number of final outputs is $N^{(n_\text{final})}=8$, and the final layer is a linear layer with complex weights.
As above, we use progressive learning.
Approximating the correlation matrix with a single gauge configuration is sufficient.
After training, we select the best networks, freeze the weights, and perform the final measurement of the correlation matrix over the whole ensemble.

\begin{figure}
    \centering
    \includegraphics[width=0.45\textwidth]{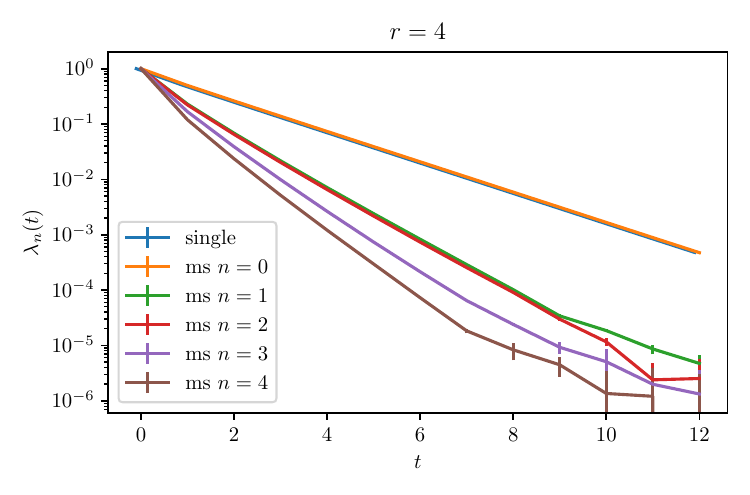}
    \includegraphics[width=0.45\textwidth]{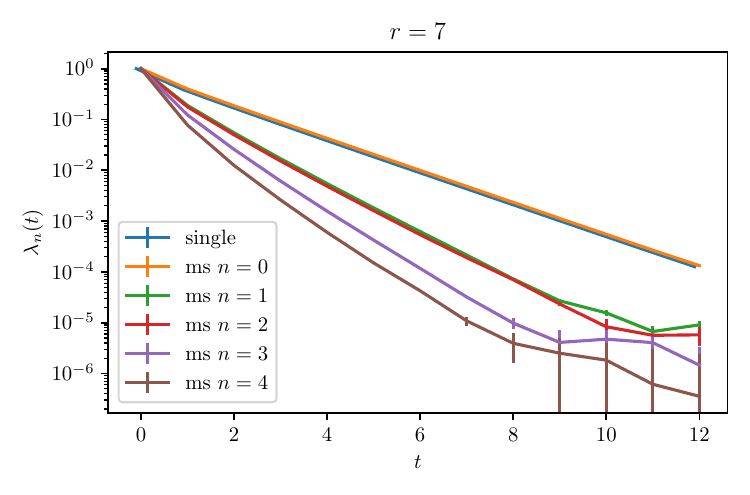}
    \caption{The generalized eigenvalue from the final measurement for $r=4$ and $r=7$ in a logarithmic plot. The data points are connected with lines for better visualization. The label ``single'' indicates the measurement for the single state network, while ``ms $n=\underline{\hspace{0.7em}}$'' indicates the actual eigenvalues from the final measurement. As expected, the ground state of the tower of states has the same mass as the single state network.}
    \label{fig:loop_ms_correlator_comparison}
\end{figure}

Figure~\ref{fig:loop_ms_correlator_comparison} shows the final correlator of the excited states, i.e., the functional form of the generalized eigenvalues in a logarithmic scale for two separations $r$. We observe that the ground state ($n=0$) exhibits the same mass as the state from the single-state network. This is expected because all approaches should converge to the same ground state. Furthermore, we see that the first two excited states ($n=1,2$) share the same mass. This is also to be expected since the next-highest states are in the doublet~$\Pi_u$.
The third state ($n=3$) has a different mass than the states below and above, which is again anticipated, since this state should be the singlet~$\Sigma_u^-$. Overall, we conclude that this method effectively generates interpolators for a set of optimized excited states in an automated manner. This is a significant achievement, considering that constructing a set of interpolators is typically a challenging task.

\begin{figure}
    \centering
    \includegraphics[width=0.45\textwidth]{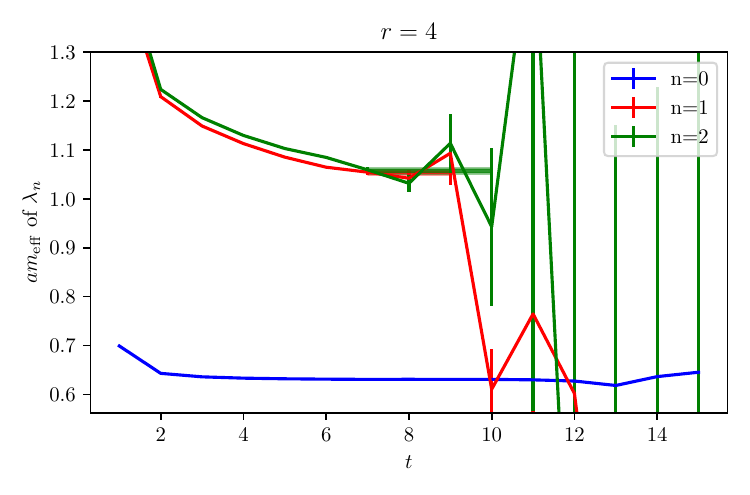}
    \includegraphics[width=0.45\textwidth]{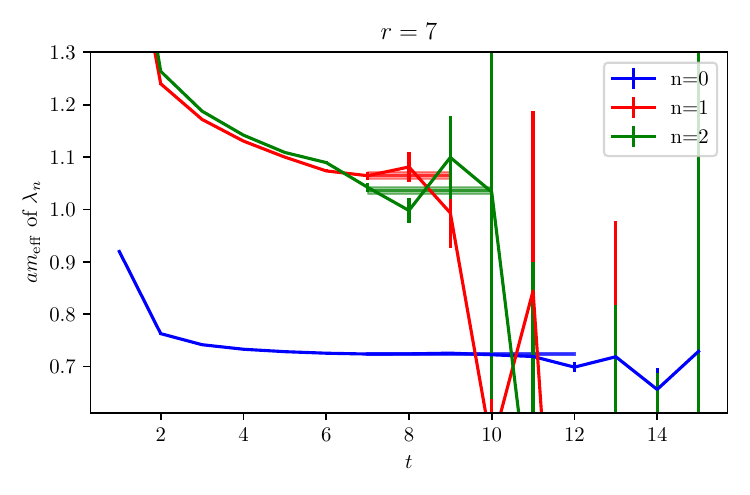}
    \caption{The effective masses of the first three states for $r=4$ and $r=7$.}
    \label{fig:meff_ms_comparison}
\end{figure}

Figure~\ref{fig:meff_ms_comparison} shows the effective masses of the first three states.
The first (ground state) exhibits a clear plateau; the first two excited states converge to a common plateau, although the signal is lost for larger times.
We extract the plateaus by constant fits in a reasonable fit range.
\begin{figure}
    \centering
    \includegraphics[width=0.45\textwidth]{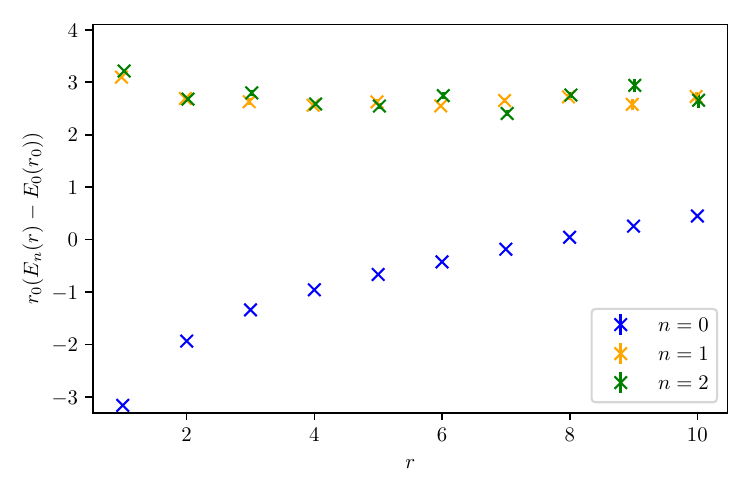}
    \caption{The static energies of the first three states as a function of the distance $r$.}
    \label{fig:static_energies_excited_states}
\end{figure}

Figure~\ref{fig:static_energies_excited_states} shows the $r$~dependence of the first three states' static energies.
We see that the static energies of the first two excited states do not exhibit a Coulomb-like ($\propto 1/r$) behavior. Instead, they follow a slightly repulsive behavior. For larger $r$, the $r$ dependence is milder. This shape of the functional curve is exactly what we would expect for the static energies of $\Pi_u$, see, for example, Fig.~2 of Ref.~\cite{Schlosser:2021wnr}.
It is the octet repulsion predicted for the hybrid static energies in Born-Oppenheimer effective field theory~\cite{Brambilla:1999xf, Berwein:2015vca}.

The evidence presented in Figs.~\ref{fig:loop_ms_correlator_comparison}--\ref{fig:static_energies_excited_states} strongly suggests that the first excited state corresponds to the $\Pi_u$ state, while the ground state corresponds to the $\Sigma_g^+$ state.
Because we have not specified the quantum numbers explicitly, it is possible that the network could have identified an unanticipated state between $\Sigma_g^+$ and $\Pi_u$.
This possibility could be more definitively investigated by searching for excited states at off-axis separations and cross-checking the energy levels from all orientations, thereby assembling the $\Pi_u$ doublet and, perhaps, other multiplets.
Constructing off-axis interpolators will be a challenge, as the cubic group theory for off-axis separations is more complicated and has not yet been addressed.

In conclusion, we have demonstrated that the neural-network approach can effectively identify excited states without requiring additional constraints. This discovery opens up new possibilities for future applications, especially in cases where the system at hand is not yet fully understood.

\subsection{Unveiling the networks}
\label{sec:unveiling_network}

The final network consists of a superposition of differently shaped paths connecting the starting and end points, with each path weighted by an optimized coefficient.
The actions of the layers and, hence, of the network are clearly defined and can be used to algebraically trace the learned structure of the neural network. In this section, we introduce a formalism to describe and then \emph{visualize} the neural network's superposition of three-dimensional paths.

The path of a straight Wilson line, along with all possible plaquette insertions, can be explicitly represented in algebraic form.
For instance, if $r=3$ and $\hat{\mathbf{e}}_r=\hat{\mathbf{e}}_x$, the straight line might be represented as $(x,x,x)$, where each element in the sequence corresponds to a step in the directions $\pm x$, $\pm y$, or $\pm z$.
Likewise, an example for a  plaquette-inserted Wilson line is $(y,x,-y,x,x)$ which corresponds to the plaquette insertion $\phi_{0,y,x}$ as defined in Eq.~\eqref{eq:plaquette_insertion_definition}.
By following this method, we can derive an algebraic representation for each initial element.

A sum of paths, as created in the linear layer, can be represented as a sequence of those paths, each with an assigned weight.
A sequence of paths $p_i$ with weights $w_i$ can be written as $[(w_1,p_1),(w_2,p_2),\ldots]$.
Adding a second superposition of the same paths, now with weights $w_i'$, is given by $[(w_1,p_1),\ldots]+[(w_1',p_1),\ldots]=[(w_1+w_1',p_1),\ldots]$.
This prescription enables us to analytically trace the action of the linear layer by following the definition in Eq.~\eqref{eq:linear_layer}, and, thus, it can be used to extract the learned structure of the neural network.

The convolutional layer modifies the path directly. For instance, the introduced straight line and the plaquette insertion changes by a convolution in $z$ direction into $(z,x,x,x,-z)$ and $(z,y,x,-y,x,x,-z)$, respectively. The new paths are assigned a weight and summed up with a set of other convoluted paths. Therefore, the actions of the convolutional layer, as defined in Eq.~\eqref{eq:convolutional_layer_analytical_definition}, can also be algebraically traced and used to extract the learned structure of the neural network.

We can establish similar rules for the bilinear layer.
Here, however, we focus exclusively on convolutional and linear layer networks for reasons mentioned below.

Overall, the neural network effectively maps the set of straight Wilson lines and their plaquette-inserted modifications into a sum of less-trivially shaped Wilson lines represented as an algebraic sequence of weights and the corresponding algebraic form of the path. These weights are dictated by the neural network's parameters, and we obtain the optimized set after training the network.

We focus on a neural-network architecture consisting only of linear and a few convolutional layers. With bilinear layers and numerous convolutional layers, we found that storing all weights and their corresponding paths becomes memory demanding, particularly for networks that contain bilinear layers.
Therefore, for visualization we use neural networks with one linear and three convolutional layers each. These networks are less optimal in terms of ground-state overlap than the more complex ones used in the rest of this study.
They should be good enough to provide a first impression of how the neural network learns an optimal superposition of shapes.

\begin{figure}
    \centering
    \includegraphics[width=0.4\textwidth]{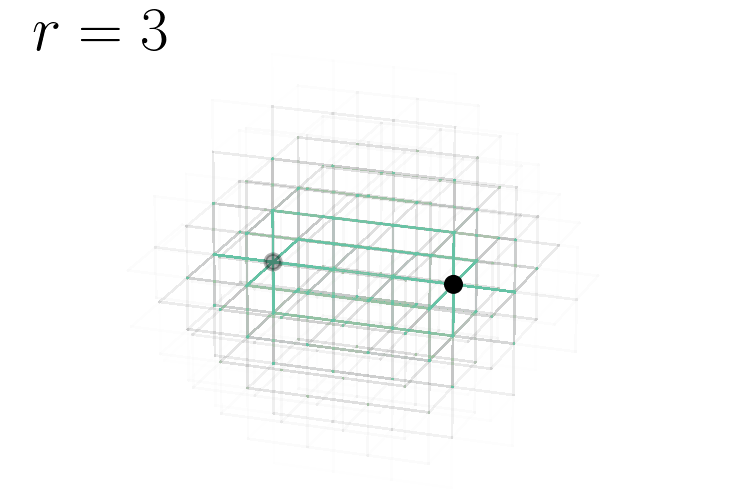}
    \includegraphics[width=0.4\textwidth]{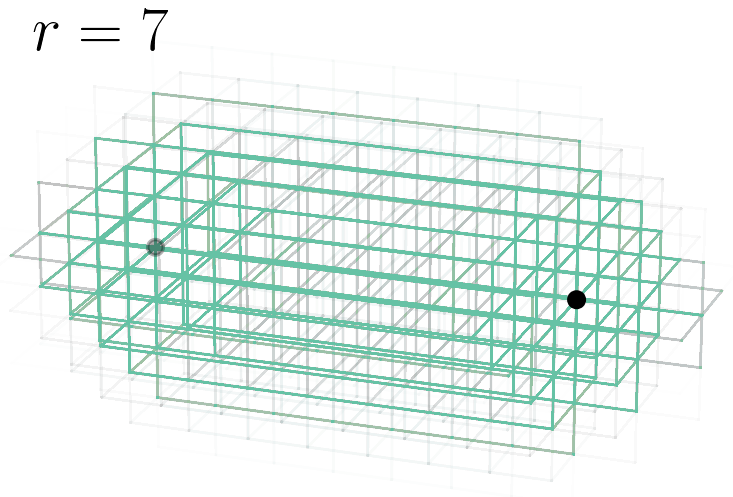}
    \caption{Three-dimensional visualization of the neural-network-optimized operator for the ground state of a quark-antiquark pair at particle separations $r=3$ and $r=7$. The black dots mark the positions of the sink and source of the static pair, and the turquoise lines represent the Wilson lines that are present in the optimized operator.
    The opacity of each line is proportional to the absolute value of the weight in the optimized operator.}
    \label{fig:path_visualization_ground_state}
\end{figure}

By tracing the analytical operations of the layers using the optimized values of the neural network's weights, we compute the weight of each path that appears in the optimized operator for the static quark-antiquark ground state.
We then visualize these paths in a three-dimensional plot, with the opacity of each path being proportional to the absolute value of its weights.

The resulting picture is shown in Fig.~\ref{fig:path_visualization_ground_state} for two quark-antiquark separations, $r=3$ and $r=7$.
In both cases, the neural-network-optimized Wilson line is several lattice spacings thick.
The contribution of link variables for paths that are farther away from the central straight line decreases gradually, indicating a radial decay of the gluon dynamics.
The learned structure resembles a flux tube, where gluonic energy is concentrated in a cylinder around the separation axis.
Additionally, the flux tube appears to have a rotational symmetry around the separation axis and a reflection symmetry around the center of the separation.
This is expected since the ground state $\Sigma_g^+$ is defined by those quantum numbers.

Now we repeat the procedure for the excited state. At this stage, we focus on a neural network with one linear layer and one convolutional layer, as deeper networks were too memory demanding. A more efficient algorithm is to be developed in future studies.

In contrast to the single-state network, the excited-state neural network has multiple outputs, where the correct sum of those outputs generates the corresponding state. The correct weights for each output are not initially known, and a measurement over the lattice ensemble is required to perform the GEVP. We perform the measurement of the correlation matrix and use the generalized eigenvalue $v_n(t,t_0)$ as the coefficient vector for the final sum.

Nevertheless, a defining property of excited states is their quantum number, especially for $\Pi_u$, which transforms oddly under $\mathcal{P}\circ\mathcal{C}$ and each $\pi/2$ rotation is assigned with a phase shift of $e^{i\pi/2}$. Thus, the complex phase of each path plays as important a role as the magnitude. Therefore, in addition to the variation of the opacity, we employ a color code indicating the complex phase of the paths. This complicates the picture, but it still provides an impression of what the neural network has learned. For example, if the phases do not matter (as in the $\Sigma_g^+$ state), we will not be able to identify regions dominated by a specific phase since all possible phases are summed. In contrast, if the phase matters, we can identify regions of a specific phase since paths with a comparable phase, and hence, a similar color, are summed.

\begin{figure*}
    \centering
    \includegraphics[width=0.9\linewidth]{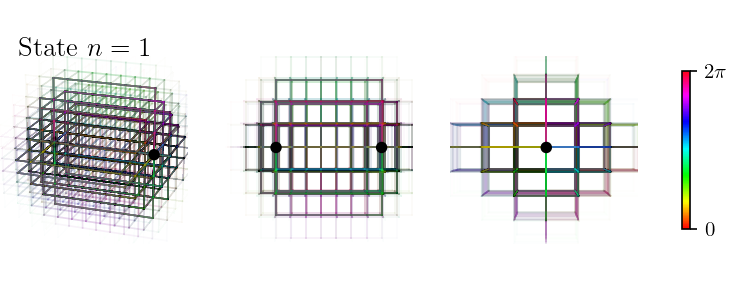}
    \includegraphics[width=0.9\linewidth]{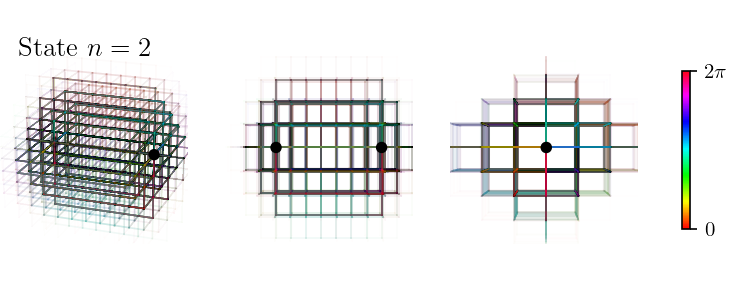}
    \caption{Visualizations of the neural-network-optimized operator for the first two excited states (top, $n=1$; bottom, $n=2$) of the quark-antiquark pair for a particle separation of $r=7$, shown from three different perspectives. The paths are color coded by the complex phases of their coefficients in the final sum, and the opacity is proportional to the absolute value of these coefficients.}
    \label{fig:path_visualization_state1}
\end{figure*}

Figure~\ref{fig:path_visualization_state1} illustrates the cumulative paths of the first two excited states from three different perspectives: The first view is from an off-axis side angle, the second is an on-axis view orthogonal to the separation axis, and the third perspective is from above, along the separation axis.

Considering the $n=1$ excited state, from the top view, we can clearly identify four regions where different colors predominate with continuous transitions. The right top region is dominated by green, indicating a complex phase of $\sim\pi/2$, the lower top by red, indicating a complex phase of $\sim 2\pi$/$\sim 0$, the lower left by purple, indicating a complex phase of $\sim 3\pi/2$, and the top left by turquoise or blue, indicating a complex phase of $\sim\pi$. The side views show that the phase is preserved over the full separations, demonstrating that it is not randomly distributed.

These observations demonstrate that the phase shifts accompanying $\pi/2$ rotations are approximately $e^{i\pi/2}$, as expected for $\Pi_u$ states. Paths that are reflected at the separation axis exhibit a phase shift of $-1$, which corresponds to an odd quantum number for the ($\mathcal{P} \circ \mathcal{C}$) transformation. This is the same observation we had in Fig.~\ref{fig:Pi_u_picture} for the $\Pi_u$ state, apart from a global phase difference ambiguity.

When we compare these observations with the $n=2$ excited state, we see green and red regions interchanged by reflecting along the horizontal axis and, so, reach the same conclusion. The two states are apparently related by the parity transformation, which results in the doublet of the $\Pi_u$ state. Thus, these findings make more vivid the evidence that the neural network has automatically learned the $\Pi_u$ state.

\section{Summary, conclusions, and outlook}
\label{sec:outlook}

We have developed a new method using neural networks to enhance Wilson loop measurements in a realistic lattice QCD setup.
We have introduced gauge-equivariant layers and established a loss function that optimizes for the ground state in a Euclidean correlator.
The trained neural network performs as well as, or even better than, Coulomb-gauge Wilson lines, while providing a gauge-invariant setup.
This achievement is significant and beneficial for contexts where the Coulomb gauge is not easily applicable.

Further, we have demonstrated that neural-network-based Wilson loops can strikingly improve ground-state expectation values, such as the direct measurement of static forces.
We have also explored combining these neural networks with other algorithms, specifically by integrating the multilevel algorithm, to further enhance their application.

Finally, our neural-network method can find the tower of excited states without explicitly specifying the quantum numbers. We found that the static energies of the first excited states exhibit characteristics consistent with the $\Pi_u$ hybrid state, indicating that the network has successfully identified the $\Pi_u$ state. This result confirms that the first excited state is indeed the $\Pi_u$ hybrid state.

Overall, this study provides a foundation for future studies. Some of them might follow directly from this work, and others are extensions of this method to other systems.

A natural next step is to extend this method to off-axis separations to optimize both ground and excited states and to perform high-precision lattice-QCD calculations of, for instance, hybrid static energies and other ground-state expectation values.
The combination with multilevel has shown a promising approach to reach high precision.
Other combinations with gradient flow or HYP smearing in unquenched theories are also applicable.
Furthermore, tailored networks that obey specific quantum numbers can be developed, opening the possibilities of this method for more specialized cases in the future.

Another direction is measuring Wilson loops at non\-zero temperature.
The attainment of the effective-mass plateau at shorter times may help address the challenge of extracting the static energy from the Wilson loop~\cite{Bazavov:2023dci,Ali:2025iux}.

In addition to these points, several open questions remain. Further research should clarify the extent to which a neural network can improve the Wilson loop (especially for larger $r$), identify the specific quantum numbers it learns, and explore more effective visualization techniques to illustrate what it has learned.

In the long term, this method can be extended to the generalized Wilson loops required for Born-Oppenheimer effective field theories (BOEFTs) \cite{Berwein:2015vca,Soto:2020xpm,Berwein:2024ztx}
that describe tetra- and pentaquark systems. We believe our approach could be particularly convenient for studying these generalized Wilson loops. Notably, the exotic $XYZ$ quarkonia states have recently garnered significant attention in both lattice field theory and experimental studies. The BOEFT framework allows the direct incorporation of lattice input for static energies into phenomenological analyses, providing a simpler, alternative way to study scattering problems in lattice QCD.
Wilson-loop-type objects also show up for glueball and gluelump operators. The operators consist of local Wilson loops with specific quantum numbers, so a similar technology can be adapted for this study.

Additionally, Wilson loops also play a role in the low-energy factorization of long-distance matrix elements in quarkonium NRQCD, using strongly coupled pNRQCD~\cite{Brambilla:2021abf,Brambilla:2022ayc}.
This method may enable the first lattice-QCD calculations of quarkonium production, potentially having a significant impact on LHC experiments.

Since the overall structure of the neural network---gauge-equivariant layers, loss functions for training, and optimization on a subset of configurations---is valid for all types of Euclidean correlators, we can also explore neural-network approaches for hadronic states. This can be incorporated into current computational efforts to compute parton distribution functions, hadron vacuum polarizations, and hadron structures.

\acknowledgments

We thank Tom Magorsch, Simon Pfahler, Marcel Rodekamp, and Marc-Philipp Tillschneider for their useful discussions and inputs.
The authors gratefully acknowledge the scientific support and resources of the AI service infrastructure LRZ AI Systems provided by the Leibniz Supercomputing Centre (LRZ) of the Bavarian Academy of Sciences and Humanities (BAdW), funded by Bayerisches Staatsministerium für Wissenschaft und Kunst (StMWK).
J.M.-S. acknowledges support by the Munich Data Science Institute (MDSI) at the Technical University of Munich (TUM) via the Linde/MDSI Doctoral Fellowship program.
V.B. conducted part of this work in the MIT Center for Theoretical Physics---A Leinweber Institute, and acknowledges support from the MIT Department of Physics.
N.B. acknowledges the Advanced ERC Grant No. ERC-2023-ADG-Project EFT-XYZ.
This document was prepared by the TUMQCD Collaboration using the resources of the Fermi National Accelerator Laboratory (Fermilab), a U.S. Department of Energy, Office of Science, Office of High Energy Physics HEP User Facility.
Fermilab is managed by Fermi Forward Discovery Group, LLC, acting under Contract No.\ 89243024CSC000002.

\section*{Data Availability}

The data that support the findings of this article are not publicly available. The data are available from the authors upon reasonable request.

\bibliography{bibliography}

\end{document}